\documentclass[aps,prl,twocolumn,10pt,amsmath,amssymb,superscriptaddress]{revtex4-2}
\usepackage[a4paper,colorlinks=true,
linkcolor=blue,citecolor=blue,
pdfauthor={ },
pdftitle={ },
pdfsubject={ },
pdfkeywords={ }]{hyperref}
\usepackage{graphicx}
\usepackage{color}
\usepackage{xcolor}
\usepackage{siunitx}
\usepackage{physics}
\UseRawInputEncoding
\usepackage[columnwise]{lineno}
\DeclareSIUnit\cps{\mathrm{cps}}
\DeclareSIUnit\Molar{\textsc{M}}
\DeclareSIUnit\rpm{\mathrm{rpm}}
\DeclareSIUnit\gauss{G}
\newcommand{\revisen}[1]{{#1}}

\def\Hamil{\mathcal{H}}
\def\para{\parallel}
\newcolumntype{x}{>{$}X<{$}}

\begin{document}
	
\title{Parallel accelerated electron paramagnetic resonance spectroscopy using diamond sensors}

\author{Zhehua Huang}
\altaffiliation{These authors contributed equally to this work.}
\affiliation{CAS Key Laboratory of Microscale Magnetic Resonance and School of Physical Sciences, University of Science and Technology of China, Hefei 230026, China}
\affiliation{Anhui Province Key Laboratory of Scientific Instrument Development and Application, University of Science and Technology of China, Hefei 230026, China}

\author{Zhengze Zhao}
\altaffiliation{These authors contributed equally to this work.}
\affiliation{CAS Key Laboratory of Microscale Magnetic Resonance and School of Physical Sciences, University of Science and Technology of China, Hefei 230026, China}
\affiliation{Anhui Province Key Laboratory of Scientific Instrument Development and Application, University of Science and Technology of China, Hefei 230026, China}

\author{Fei Kong}
\email{kongfei@ustc.edu.cn}
\affiliation{CAS Key Laboratory of Microscale Magnetic Resonance and School of Physical Sciences, University of Science and Technology of China, Hefei 230026, China}
\affiliation{Anhui Province Key Laboratory of Scientific Instrument Development and Application, University of Science and Technology of China, Hefei 230026, China}
\affiliation{Hefei National Laboratory, University of Science and Technology of China, Hefei 230088, China}

\author{Zhecheng Wang}
\affiliation{CAS Key Laboratory of Microscale Magnetic Resonance and School of Physical Sciences, University of Science and Technology of China, Hefei 230026, China}
\affiliation{Anhui Province Key Laboratory of Scientific Instrument Development and Application, University of Science and Technology of China, Hefei 230026, China}
\affiliation{School of Biomedical Engineering and Suzhou Institute for Advanced Research, University of Science and Technology of China, Suzhou 215123, China}

\author{Pengju Zhao}
\affiliation{CAS Key Laboratory of Microscale Magnetic Resonance and School of Physical Sciences, University of Science and Technology of China, Hefei 230026, China}
\affiliation{Anhui Province Key Laboratory of Scientific Instrument Development and Application, University of Science and Technology of China, Hefei 230026, China}

\author{Xiangtian Gong}
\affiliation{CAS Key Laboratory of Microscale Magnetic Resonance and School of Physical Sciences, University of Science and Technology of China, Hefei 230026, China}
\affiliation{Anhui Province Key Laboratory of Scientific Instrument Development and Application, University of Science and Technology of China, Hefei 230026, China}

\author{Xiangyu Ye}
\affiliation{CAS Key Laboratory of Microscale Magnetic Resonance and School of Physical Sciences, University of Science and Technology of China, Hefei 230026, China}
\affiliation{Anhui Province Key Laboratory of Scientific Instrument Development and Application, University of Science and Technology of China, Hefei 230026, China}

\author{Ya Wang}
\affiliation{CAS Key Laboratory of Microscale Magnetic Resonance and School of Physical Sciences, University of Science and Technology of China, Hefei 230026, China}
\affiliation{Anhui Province Key Laboratory of Scientific Instrument Development and Application, University of Science and Technology of China, Hefei 230026, China}
\affiliation{Hefei National Laboratory, University of Science and Technology of China, Hefei 230088, China}

\author{Fazhan Shi}
\email{fzshi@ustc.edu.cn}
\affiliation{CAS Key Laboratory of Microscale Magnetic Resonance and School of Physical Sciences, University of Science and Technology of China, Hefei 230026, China}
\affiliation{Anhui Province Key Laboratory of Scientific Instrument Development and Application, University of Science and Technology of China, Hefei 230026, China}
\affiliation{Hefei National Laboratory, University of Science and Technology of China, Hefei 230088, China}
\affiliation{School of Biomedical Engineering and Suzhou Institute for Advanced Research, University of Science and Technology of China, Suzhou 215123, China}

\author{Jiangfeng Du}
\affiliation{CAS Key Laboratory of Microscale Magnetic Resonance and School of Physical Sciences, University of Science and Technology of China, Hefei 230026, China}
\affiliation{Anhui Province Key Laboratory of Scientific Instrument Development and Application, University of Science and Technology of China, Hefei 230026, China}
\affiliation{Hefei National Laboratory, University of Science and Technology of China, Hefei 230088, China}
\affiliation{Institute of Quantum Sensing and School of Physics, Zhejiang University, Hangzhou 310027, China}

\begin{abstract}
	The nitrogen-vacancy (NV) center can serve as a magnetic sensor for electron paramagnetic resonance (EPR) measurements. Benefiting from its atomic size, the diamond chip can integrate a tremendous amount of NV centers to improve the magnetic-field sensitivity. However, EPR spectroscopy using NV ensembles is less efficient due to inhomogeneities in both sensors and targets. Spectral line broadening induced by ensemble averaging is even detrimental to spectroscopy. Here we show a kind of cross-relaxation EPR spectroscopy at zero field, where the sensor is tuned by an amplitude-modulated control field to match the target. The modulation makes detection robust to the sensor’s inhomogeneity, while zero-field EPR is naturally robust to the target’s inhomogeneity. We demonstrate an efficient EPR measurement on an ensemble of $\sim$ 30000 NV centers. Our method shows the ability to not only acquire unambiguous EPR spectra of free radicals, but also monitor their spectroscopic dynamics in real time.
\end{abstract}

\maketitle


Quantum sensing can substantially improve the measurement of physical quantities such as magnetic fields \cite{Budker2007}, electric fields \cite{Facon2016}, temperature \cite{Kucsko2013}, pressure \cite{Doherty2014}, and so on. Among the numerous quantum sensing platforms \cite{Degen2017}, the nitrogen-vacancy (NV) center in diamond is favored for its atomic size and thus nanoscale resolution \cite{Maze2008,Balasubramanian2008}. As a magnetometer, the NV sensor has realized magnetic detection at nanometer scale \cite{Du2024}, promoting electron paramagnetic resonance (EPR) to the single-spin level \cite{Shi2015,Schlipf2017,Shi2018,Pinto2020}. The trade-off is relatively low sensitivity of individual shallow NV sensors, making single-spin EPR measurements time-consuming. On the other hand, the atomic size of the NV centers allows an integration of a large number $n_{\text{NV}}$ of NVs in a limited volume with sensitivity scaling as $n_{\text{NV}}^{-1/2}$ \cite{Barry2020}. For example, NV ensemble-based magnetometers in various frequency bands have been developed to improve the field sensitivity to pT$\cdot$Hz$^{-1/2}$ level \cite{Wolf2015,Zhou2020,Zhang2021,Wang2022}. However, for NV-based EPR detection, where the signal typically comes from statistical fluctuations in the spin-generated magnetic field rather than the field itself \cite{Grinolds2013,Shi2015,Schlipf2017,Shi2018,Pinto2020}, this parallel acceleration capability has not been fully exploited.

Two key challenges have limited the utilization of NV ensembles for EPR spectroscopy. First, the statistical fluctuation signal has strong $\sim r^{-6}$ dependence on the NV-target distance $r$. 
Detection is only possible when the NV sensor and the target spin are a few ten nanometers away from each other, which means both of them are close to the diamond surface. 
Compared to the NV ensemble-based magnetometers \cite{Wolf2015,Zhou2020,Zhang2021,Wang2022}, integrating the same number of NVs in such a two-dimensional space requires a much larger area, where precise spin controls will fail due to inhomogeneous control fields. Second, different molecules typically have different orientations at the diamond-sample interfaces \cite{Schlipf2017}. 
Anisotropic interactions within paramagnetic targets will lead to orientation-dependent resonance frequencies. \revisen{Therefore, a direct summation of the EPR signals will broaden the spectrum with signal magnitude growing inefficiently. Previous studies have circumvented these challenges by utilizing NV relaxometry to detect transition metal ions \cite{Simpson2017,Grant2023,iyer_2024}. The EPR spectrum of each ion is sufficiently broad to mask their heterogeneity and enable direct signal summation. However, the problem remains in the detection of stable free radicals, such as nitroxide radicals, which are widely used in spin-label EPR to study the conformation and dynamics of proteins \cite{Borbat2001}. Extracting this molecular information usually requires high-resolution spectroscopy.}

\begin{figure*}[htbp]
\centering \includegraphics[width=\textwidth]{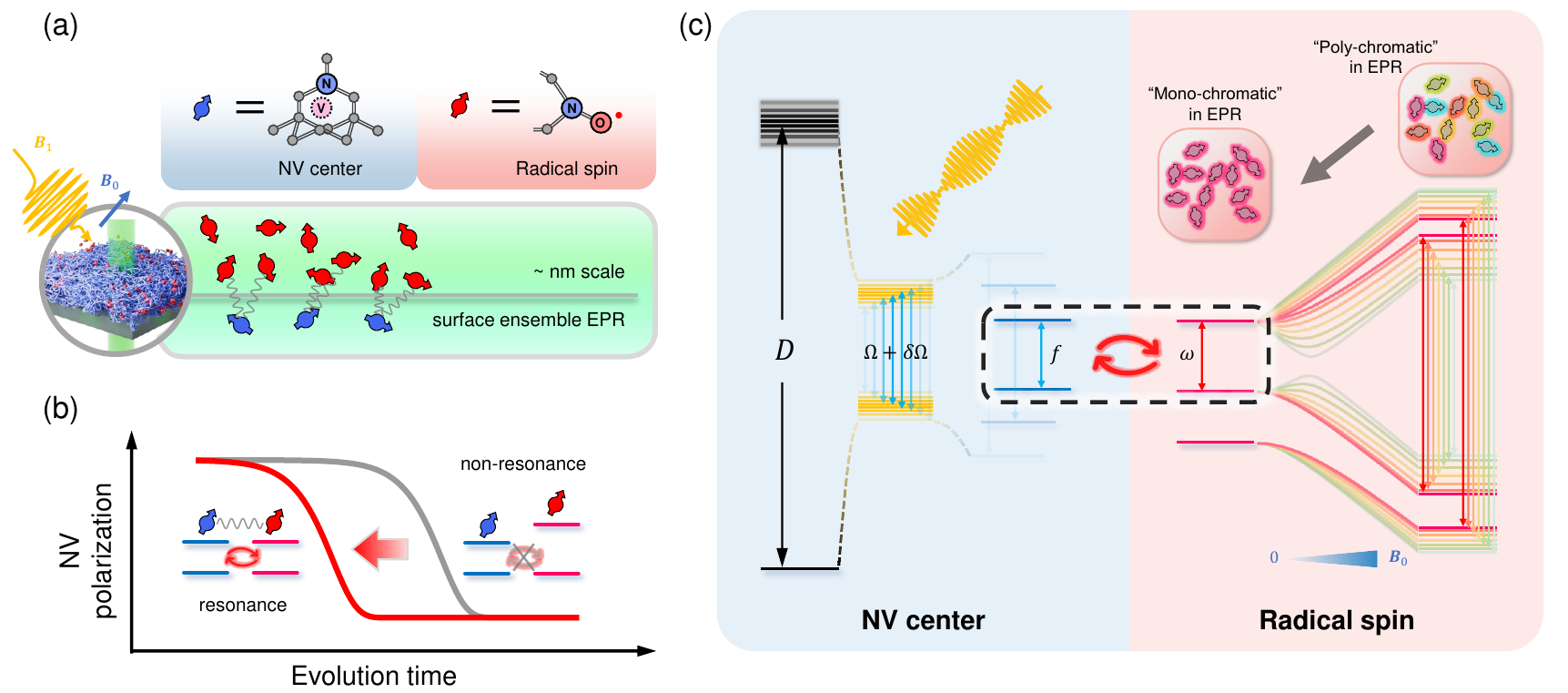} \protect\caption{
		\textbf{The parallel acceleration detection scheme.}
		(a) An illustrative diagram of NV center (blue arrow) detecting electron spins (red arrow). 
		Upper panel: The structure of NV center and nitroxide radical spin.
		Lower panel: Spectroscopic measurements necessitate the application of external fields, such as optical pulses, bias magnetic field, and control fields. In widefield setup, NV centers measure an ensemble of target radicals on surface in nano-meter scale distances.\label{subfig:figure1a}
		(b) Schematic diagram of cross-relaxation detection. Depending on whether the energy levels of the sensor and target spins match, the relaxation of the sensor NV will be accelerated or remain unchanged.
		(c) Schematic diagram of cross-relaxation based EPR spectroscopy. Left Panel: By applying an amplitude-modulated microwave sequence, all sensor spins are synchronized to have the same energy levels determined by the modulation frequency $f$. 
		Right Panel: 
		Different colored energy levels indicate different orientations of the targets, corresponding to the `emission' of different colors of light.
		By conducting experiments at zero field, the energy levels of all target spins are equivalent regardless of orientations.
		The modulation frequency is then swept to obtain spectra.
	}
	\label{fig:DetectionScheme}
\end{figure*}

\begin{figure*}[ht]
\centering \includegraphics[width=1\textwidth]{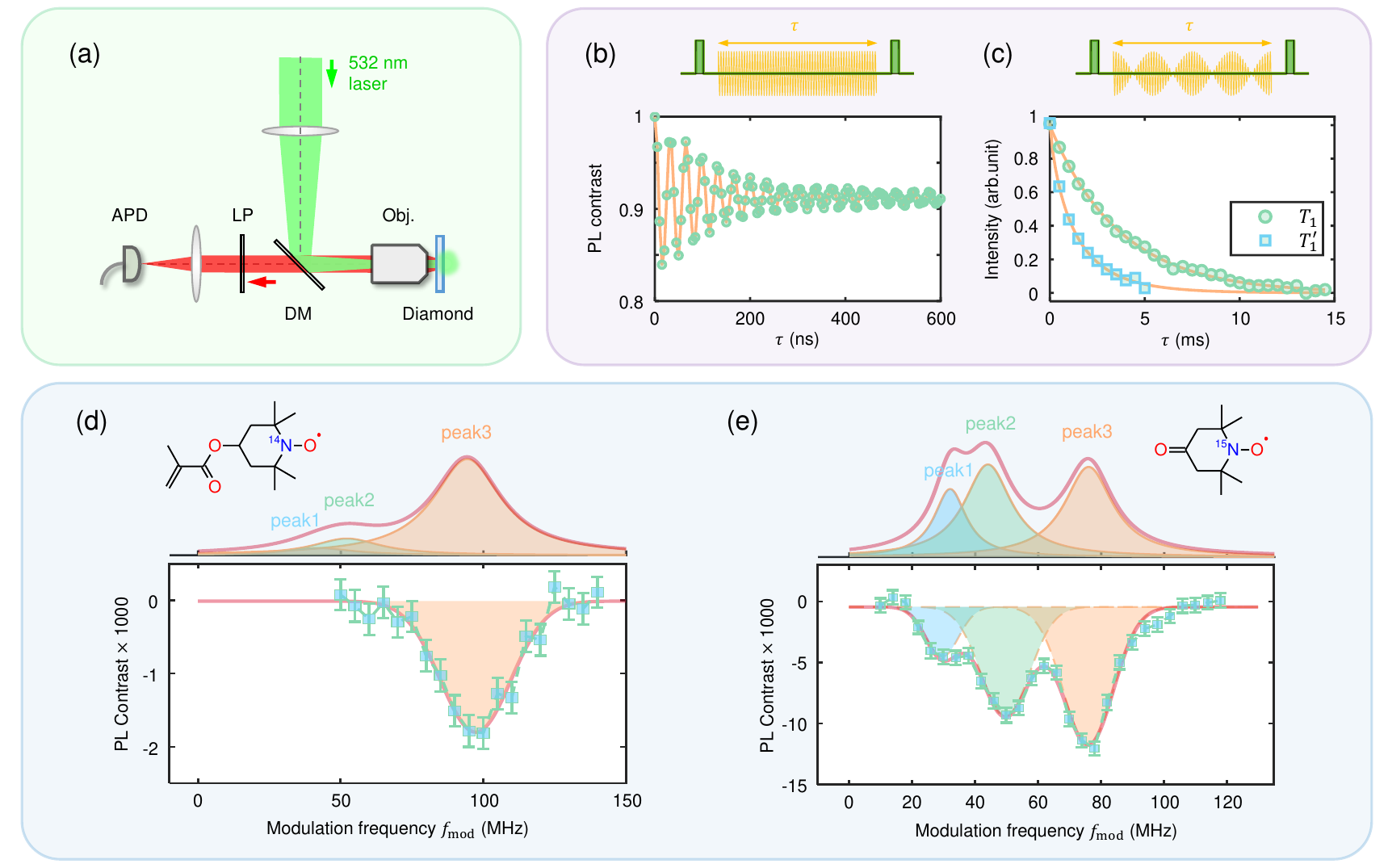} \protect\caption{
		\textbf{Experimental demonstration of parallel measurements.}
		(a) Experimental setup of the widefield microscopy. (Abbreviations: APD: Avalanche photodiodes, LP: Long pass filter, DM: Dichroic mirror, Obj.: Microscope objective).
		(b) Rabi oscillation of NV ensembles. The points are experimental results, while the line is a cosine damping fit.
		(c) Relaxation measurements. The inset gives the pulse sequence, where the polarization time and readout window are $1.04$ and $\SI{0.52}{\ms}$, respectively. The points are experimental results, while the lines are exponential fits giving $T_1=\SI{3.7\pm0.1}{\milli\second}$, $T_1^{'}=\SI{1.3\pm0.5}{\milli\second}$. The driving strength and modulation frequency are $35$ and $\SI{50}{\MHz}$, respectively.
		(d) NV-EPR measurement of $^{14}$N-TEMPO methacrylate (inset: chemical structure). The points are experimental results, while the line is a Gaussian fit. The spectrum in the upper panel is a prediction according to the conventional EPR measurements. Other measurement parameters are: Evolution time $\SI{0.9}{\milli\s}$, polarization time $\SI{0.41}{\ms}$, readout time $\SI{0.2}{\ms}$, $\kappa = 0.57$, and the laser power density  $\SI{101}{\watt\per\cm\squared}$.
		(e) NV-EPR measurement of $\text{4-Oxo-TEMPO-d}_{16}$, $^{15}$N (inset: chemical structure). The points are experimental results, while the line is a 3-peak Gaussian fit. The spectrum in the upper panel is a prediction according to the conventional EPR measurements. Other measurement parameters are: Evolution time $\SI{1}{\milli\s}$, polarization time $\SI{0.76}{\ms}$, readout time $\SI{0.25}{\ms}$, $\kappa = 0.71$, and the laser power density is $\SI{81}{\watt\per\cm\squared}$.
	}
	\label{fig:Setup_TEMPOSpec}
\end{figure*}

\begin{figure*}[htp]
\centering\includegraphics[width=1\textwidth]{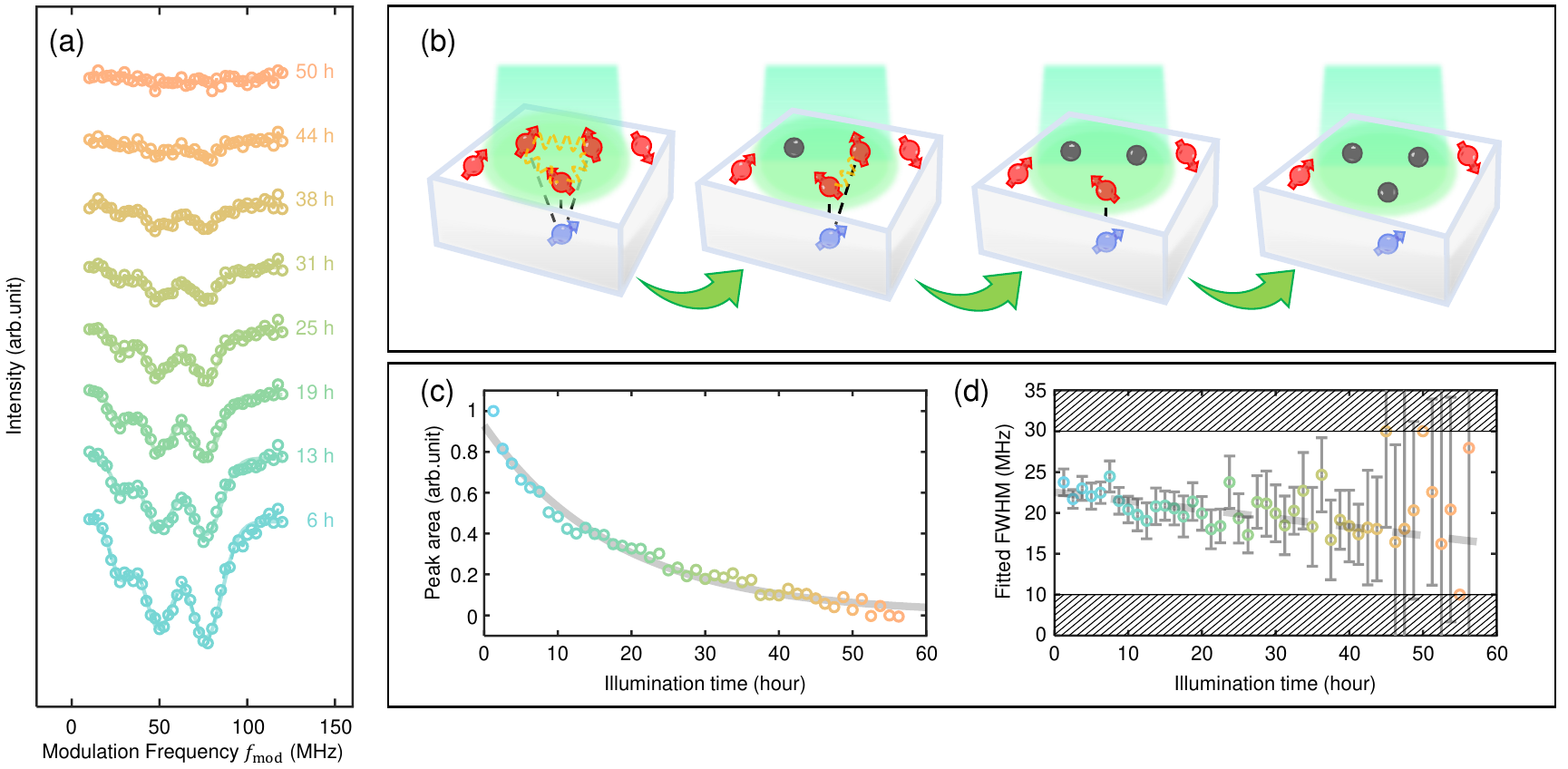}\protect\caption{
		\textbf{Dynamics of nitroxide EPR spectra}
		(a) Real-time monitoring of the spectral lines, showing a decaying trend. The points are experimental results, while the lines are 3-peak Gaussian fits.
		(Evolution time $\SI{1.2}{\milli\s}$, polarization time $\SI{1.01}{\ms}$, readout time $\SI{0.5}{\ms}$, $\kappa = 0.71$, and the laser power density is $\SI{37}{\watt\per\cm\squared}$).
		(b) Schematic diagram of the possible laser-quenching process. Gray points indicate loss of spin signal.
		(c) Dependence of the fitted peak area on total illumination time. Gray line is a further exponential fit (\revisen{$y = A\exp(-t/t_d)+y_0$}) to the fitted data (characteristic time $t_d = \SI{18\pm1}{\hour}$).
		(d) Dependence of the fitted linewidth of peak 3 on total illumination time. The error bars are the fitting errors. The hatched area represents the upper and lower limits of the fit to avoid divergence. The gray dashed line is a guide to the eye.
	}
	\label{fig:bleached_spec}
\end{figure*}

\begin{figure}[ht]
\centering \includegraphics[width=1\columnwidth]{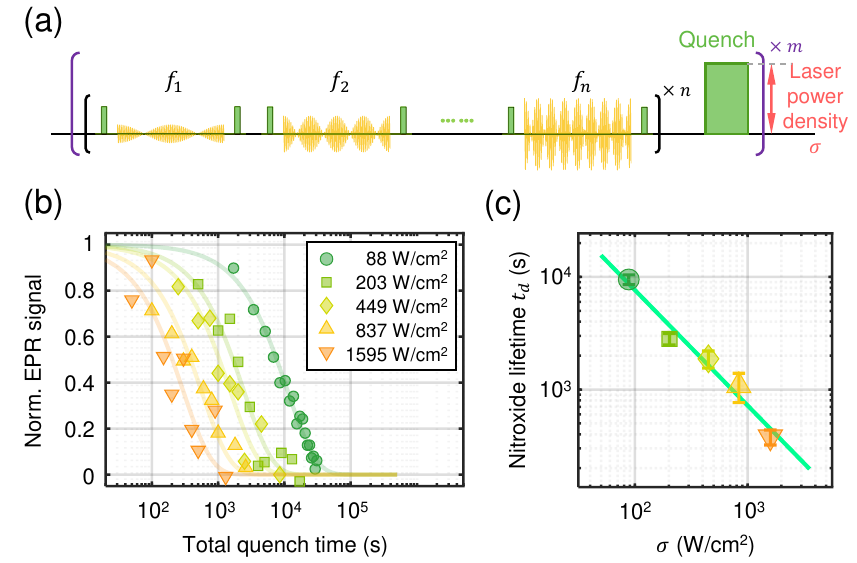}\protect\caption{
		\textbf{Dependence of quenching process on laser power density.}
		(a) Sequence of quenching experiment under high-power laser. The power of the laser quenching block($\sigma$) applied between each sequential measurement is fixed. 
		(b) 
		Quenching of EPR signal of $^{14}$N-TEMPO methacrylate under different laser power density. Data points are fitted peak areas of the sequential EPR spectra, while the lines are exponential fits, where the characteristic time is defined as the nitroxide lifetime.
		(c) Dependence of nitroxide lifetime on laser power density. Points are the mean lifetime of repeated quenching measurements with error bars indicating standard error of the mean. Line is a fit with formula of  $\log{t_d} = a\log{\sigma} + b$, which gives $a=-1.0(2),b=14(1)$, indicating a linear dependence of the quenching rate ($1/t_d$) on laser power density. 
	}
	\label{fig:SpinDecayRate}
\end{figure}

Here, we address these challenges by: (i) performing EPR measurements in the absence of external magnetic field, where the resonance frequencies of target spins are no longer orientation dependent \cite{Kong2018}; and (ii)  employing an amplitude-modulated driving field on NV ensembles, which makes EPR measurements robust to the control field inhomogeneity \cite{Qin2023}. We demonstrate a parallel-accelerated EPR spectrometer on an ensemble of $n_{\text{NV}} \sim 3\times 10^4 $ NV centers within an area of $\SI{2.3e-4}{\mm\squared}$, which can acquire clear characteristic EPR spectra of nitroxide radicals with different isotopes. Benefiting from the improved measurement efficiency, we are able to monitor the signal decay process of radicals under laser illumination. Furthermore, we investigate the dependence of the decay rate on the laser power density, providing solid evidence for laser-induced radical quenching.

The NV center electron has a spin triplet ground state with a zero-field splitting of $D_\mathrm{gs}=\SI{2.87}{\GHz}$ between $\ket{m_s=0}$ state and a two-fold degenerate $\ket{m_s=\pm 1}$ state. Those states are sensitive to magnetic fields generated by other spins, and thus can be used for EPR sensing. Besides, the NV electron spin can be polarized to $\ket{m_s=0}$ by green laser and read out by its spin-dependent fluorescence. To realizing efficient parallel acceleration, the key is to make the signal an extensive quantity, whose magnitude is additive for each sensor. The photon signal of NV centers is certainly additive. So our scheme utilizes ensembles of NV centers [Fig.~\ref{fig:DetectionScheme}(a)]. But the spectroscopic signal is not necessarily additive if different NV centers get different spectra. Our scheme is to make the signals from all NV centers have the same frequency response. 

Our measurement is based on the cross-relaxation technique [Fig.~\ref{fig:DetectionScheme}(b)]. The EPR signal arises from an increase in NV relaxation rate when the energy level splittings of the NV sensor and the target match. In the absence of external magnetic field, the energy levels of the target are solely determined by its intrinsic interactions,  and thus different targets share the same transition frequencies \cite{Bogle1961,Cole1963,Bramley1983}. In order to reliably match them, we apply an amplitude-modulated continuous driving microwave (MW) $B_1 \cos (ft) \cos (D_\mathrm{gs}t)$ on the NV centers \cite{Qin2023}. For a specific target transition, this sensor-target system can be described by a simplified Hamiltonian \cite{SOM}:
\begin{equation}
	\Hamil = \frac{\kappa}{4}\mathcal{D}_{z\perp}S_x T_x + (\omega-f) T_z,
\end{equation}
where $\mathbf{S}$ is the spin operator of the NV sensor, $\mathbf{T}$ is the reduced spin operator of the target corresponding to the specific target transition,
$\mathcal{D}_{z\perp} = \sqrt{\mathcal{D}_{zx}^2+\mathcal{D}_{zy}^2}$ \revisen{is the transverse component of dipole-dipole coupling depending} on the relative position and orientation of the target with respect to the NV sensor and the specific target transition,
\revisen{$f$ is the modulation frequency, $\omega$ is an allowed transition frequency of target spin,}
and $\kappa = \gamma_\textrm{NV} B_1/f$ is the relative driving index depending on the MW amplitude $B_1$ with $\gamma_\textrm{NV}$ being the gyromagnetic ratio of NV center.
A resonant cross relaxation happens at $f=\omega$ with the rate determined by $(\kappa\times\mathcal{D}_{z\perp})$. Therefore, the sensor or sample inhomogeneity will only affect EPR line magnitudes rather than line positions [Fig.~\ref{fig:DetectionScheme}(c)]. 

We perform the experimental demonstrations on a home-built widefield microscope, where the camera is replaced by a single photon counter to lower the readout noise [Fig.~\ref{fig:Setup_TEMPOSpec}(a)]. The density of near-surface NV centers is $\SI{1.2e3}{\textrm{NV}\per\micro\meter\squared}$ with a mean depth of $\sim\SI{8}{\nm}$.  
There are $\sim 3\times 10^4$ NVs in a circle view with $\sim \SI{17}{\micro\meter}$ diameter. \revisen{Since the detection range of NV center is comparable to the depth, the total sensing volume is $\sim \SI{1.8}{\femto\L}$.}
To ensure the total photon count rate is within the dynamical range of the counter, the laser power is much lower than the saturation power, leading to a long polarization and readout time. Fortunately, it does not sacrifice the sensitivity of relaxometry because the signal accumulation time is on the order of $T_1 \sim \SI{}{\ms}$. The targets are TEMPO molecules dispersed in polymethyl methacrylate (PMMA) and covered on diamond surface via spin coating, \revisen{with a equivalent concentration of $\SI{100}{\milli\Molar}$ \cite{SOM}. There are $\sim 1 \times 10^8$ molecules in the sensing volume.}

By applying a resonant MW of frequency $\SI{2870}{\MHz}$, the photon signal shows a normal Rabi oscillation, which decays quickly due to the driving-field inhomogeneity [Fig.~\ref{fig:Setup_TEMPOSpec}(b)]. By applying an additional amplitude modulation of frequency $\SI{50}{\MHz}$, the oscillation turns to an exponential decay with characteristic time of $T_1^{'}=\SI{1.3\pm0.5}{\milli\second}$ [Fig.~\ref{fig:Setup_TEMPOSpec}(c)]. We note that $T_1^\prime$ positively depends on the modulation frequency and negatively depends on the driving strength. In order to measure the zero-field EPR spectrum, we fix the driving length, and sweep the modulation frequency $f$, while the MW amplitude $B_1$ is swept accordingly to keep the relative driving index $\kappa=\Omega/f$ as a constant. 

As shown in Fig.~\ref{fig:Setup_TEMPOSpec}(d), a clear peak appears at $f=\SI{97\pm1}{\mega\hertz}$, consistent with the strongest resonance transition for $^{14}\mathrm{N}$ nitroxide (see \cite{SOM} Note 2, \revisen{three spectral peak positions are: $\omega_1=\frac{3}{4}A_\para-\frac{1}{4}\sqrt{8A_\perp^2 + A_\para^2}$, $\omega_2 = \frac{1}{2}\sqrt{8A_\perp^2 + A_\para^2}$, $\omega_3 =\frac{3}{4}A_\para+\frac{1}{4}\sqrt{8A_\perp^2 + A_\para^2}$. Substituting the hyperfine coupling constants $[A_\perp,A_\para] = [19,91]\,\SI{}{\MHz}$ extracted from conventional EPR results yields predicted peak positions of: $[\omega_1,\omega_2,\omega_3] = [42,53,95]\,\SI{}{\MHz}$.}). The other two peaks at $\sim$ 41 and \SI{55}{MHz} are too weak to be resolved from the baseline. We note the baseline is calibrated by subtracting a blank spectrum, which is acquired at an area without samples (\revisen{see \cite{SOM} Note 6}) or the same area after laser-quenching of the nitroxide radicals (see \cite{SOM} Note 5). To further confirm the spectrum, we repeat the measurement on $^{15}\mathrm{N}$ TEMPO molecules. As shown in Fig.~\ref{fig:Setup_TEMPOSpec}(e), three peaks appear at $27(1)$, $48.2(8)$, and $\SI{76.7\pm0.7}{\MHz}$, again consistent with the expectations (see \cite{SOM} Note 2, \revisen{$\omega_1 = A_\perp$, $\omega_2 = \frac{1}{2}(A_\para-A_\perp)$, $\omega_3 = \frac{1}{2}(A_\para+A_\perp)$. Substituting $[A_\perp,A_\para] = [32,120]\,\SI{}{\MHz}$ yields: $[\omega_1,\omega_2,\omega_3] = [32,44,76]\,\SI{}{\MHz}$.}). 
We perform all the experiments at ambient fields without magnetic shielding. The geomagnetic field ($\sim$ 50 \textmu{}T) has negligible effect here due to the broad spectra.


The measurement in Fig.~\ref{fig:Setup_TEMPOSpec}(e) costs \SI{30}{\min} with a signal noise ratio (SNR) of 25.4, which means \SI{2.8}{\s} is enough to obtain the radical spectrum. It allow us to monitor the photochemical reactions of nitroxide radicals in real time. As shown in Fig.~\ref{fig:bleached_spec}(a), the EPR spectra change significantly during long-term measurements. The EPR signal, defined by the peak area, shows an obvious decay [Fig.~\ref{fig:bleached_spec}(c)], suggesting degradation happens on either the sensor or the sample. The latter is more likely, because the spectral linewidth is also narrowing accordingly [Fig.~\ref{fig:bleached_spec}(d)]. \revisen{The signal linewidth is the sum of the linewidths of both the sensor and the sample, where the latter is contributed by} dipole-dipole interactions among the radicals, proportional to the radical concentration \cite{Simpson2017}. We can extract a reliable lifetime of nitroxide radicals from the dynamics of the EPR signal, which shows an exponential decay with a characteristic time of about $\SI{18\pm1}{\hour}$ [Fig.~\ref{fig:bleached_spec}(c)]. 

After $\sim\SI{56}{\hour}$ of laser irradiation, the EPR signal is almost extinct and does not recover in the darkness. A similar phenomenon occurs when switching to other regions that have not been irradiated before, ruling out overall degradation of the sample. To further confirm the necessity of lasers, we investigate the dependence of this kinetic process on laser power. Since we can obtain high-SNR spectra in a short period of time, we insert a segment of continuous laser between each measurement period and change the laser power. Figure \ref{fig:SpinDecayRate}(b) shows that the EPR signal decays faster under the irradiation of higher-power laser. Moreover, the decay rate has almost linear dependence on the laser power density [Fig.~\ref{fig:SpinDecayRate}(c)].

In conclusion, we conduct zero-field EPR spectroscopy of nitroxide radicals on an ensemble of near-surface NV centers. Since our scheme is robust to the inhomogeneity of both the NV sensors and the target spins, the measurement can be performed in parallel with dramatically reduced time consumption. 
We achieve an SNR of 25 with 30 measurement points within \SI{30}{\min}, corresponding to a minimum measurement time of \SI{0.1}{\s} per point.
This acceleration allows direct observation of the laser quenching process of radicals.

Our method provides a powerful tool for investigating the chemical reaction kinetics of radicals under high-power laser irradiation, which is hardly accessed by conventional EPR spectrometers. For example, the EPR sample tube typically has a diameter of $\sim$ 3 mm. To achieve the laser power density in Fig.~\ref{fig:SpinDecayRate}, the laser power should be $6\sim\SI{110}{\W}$, and then the laser-heating effect will be unacceptable. A direct application of this tool is to study the laser quenching mechanism of \revisen{spin labels such as} nitroxide radicals. Clarifying the mechanism is critical to prolonging the lifetime of spin labels, and thus to removing the major obstacle for single-molecule EPR applications at ambient conditions \cite{Schlipf2017,Shi2018}. Similarly, our method can be generalized to other applications where the sample volume is limited, such as measurements of paramagnetic defects in 2D materials \cite{kianinia_quantum_2022}.

The measurement efficiency can be further improved by utilizing more shallow NV centers. The current limit on available NVs comes from the saturation count rate of the photodetector we use , which is $\sim$ 10 Mcps. The relaxation measurement in this work uses a laser excitation power as weak as $1/3700$ of the saturation power, where the photon count rate of a single NV center is \SI{120}{\cps}. So the available NV number is 85000. The use of photodetectors with higher saturation power is a direct way to eliminate this limitation, but may introduce more readout noise than the photon shot noise \cite{Wang2022}. Another way to boost the measurement is to improve the photon collection efficiency, such as by using a parabolic-shaped lens \cite{Wolf2015} or a total internal reflection lens \cite{Xu2019}. It is also helpful for reducing the ratio of extra readout noise. 
\revisen{Moreover, the utilization of a larger size of NV ensemble can reduce the spatial heterogeneity in  EPR signal, making it a quantitative tool to determine the spin density of paramagnetic samples.}

This work was supported by the National Natural Science Foundation of China (grant nos. T2125011), the CAS (grant nos. GJJSTD20200001, YSBR-068), Innovation Program for Quantum Science and Technology (Grant No. 2021ZD0302200, 2021ZD0303204), New Cornerstone Science Foundation through the XPLORERPRIZE, and the Fundamental Research Funds for the Central Universities. 
\renewcommand\refname{Reference}


\begin{thebibliography}{10}
\expandafter\ifx\csname url\endcsname\relax
  \def\url#1{\texttt{#1}}\fi
\expandafter\ifx\csname urlprefix\endcsname\relax\def\urlprefix{URL }\fi
\providecommand{\bibinfo}[2]{#2}
\providecommand{\eprint}[2][]{\url{#2}}

\bibitem{Budker2007}
\bibinfo{author}{Budker, D.} \& \bibinfo{author}{Romalis, M.}
\newblock \bibinfo{title}{Optical magnetometry}.
\newblock \emph{\bibinfo{journal}{Nat. Phys.}} \textbf{\bibinfo{volume}{3}},
  \bibinfo{pages}{227--234} (\bibinfo{year}{2007}).

\bibitem{Facon2016}
\bibinfo{author}{Facon, A.} \emph{et~al.}
\newblock \bibinfo{title}{A sensitive electrometer based on a rydberg atom in a
  schr{\"o}dinger-cat state}.
\newblock \emph{\bibinfo{journal}{Nature}} \textbf{\bibinfo{volume}{535}},
  \bibinfo{pages}{262--265} (\bibinfo{year}{2016}).

\bibitem{Kucsko2013}
\bibinfo{author}{Kucsko, G.} \emph{et~al.}
\newblock \bibinfo{title}{Nanometre-scale thermometry in a living cell}.
\newblock \emph{\bibinfo{journal}{Nature}} \textbf{\bibinfo{volume}{500}},
  \bibinfo{pages}{54--58} (\bibinfo{year}{2013}).

\bibitem{Doherty2014}
\bibinfo{author}{Doherty, M.~W.} \emph{et~al.}
\newblock \bibinfo{title}{Electronic properties and metrology applications of
  the diamond {NV}$^{-}$ center under pressure}.
\newblock \emph{\bibinfo{journal}{Phys. Rev. Lett.}}
  \textbf{\bibinfo{volume}{112}}, \bibinfo{pages}{047601}
  (\bibinfo{year}{2014}).

\bibitem{Degen2017}
\bibinfo{author}{Degen, C.~L.}, \bibinfo{author}{Reinhard, F.} \&
  \bibinfo{author}{Cappellaro, P.}
\newblock \bibinfo{title}{Quantum sensing}.
\newblock \emph{\bibinfo{journal}{Rev. Mod. Phys.}}
  \textbf{\bibinfo{volume}{89}}, \bibinfo{pages}{035002}
  (\bibinfo{year}{2017}).

\bibitem{Maze2008}
\bibinfo{author}{Maze, J.~R.} \emph{et~al.}
\newblock \bibinfo{title}{Nanoscale magnetic sensing with an individual
  electronic spin in diamond}.
\newblock \emph{\bibinfo{journal}{Nature}} \textbf{\bibinfo{volume}{455}},
  \bibinfo{pages}{644--647} (\bibinfo{year}{2008}).

\bibitem{Balasubramanian2008}
\bibinfo{author}{Balasubramanian, G.} \emph{et~al.}
\newblock \bibinfo{title}{Nanoscale imaging magnetometry with diamond spins
  under ambient conditions}.
\newblock \emph{\bibinfo{journal}{Nature}} \textbf{\bibinfo{volume}{455}},
  \bibinfo{pages}{648--651} (\bibinfo{year}{2008}).

\bibitem{Du2024}
\bibinfo{author}{Du, J.}, \bibinfo{author}{Shi, F.}, \bibinfo{author}{Kong,
  X.}, \bibinfo{author}{Jelezko, F.} \& \bibinfo{author}{Wrachtrup, J.}
\newblock \bibinfo{title}{Single-molecule scale magnetic resonance spectroscopy
  using quantum diamond sensors}.
\newblock \emph{\bibinfo{journal}{Rev. Mod. Phys.}}
  \textbf{\bibinfo{volume}{96}}, \bibinfo{pages}{025001}
  (\bibinfo{year}{2024}).

\bibitem{Shi2015}
\bibinfo{author}{Shi, F.} \emph{et~al.}
\newblock \bibinfo{title}{Single-protein spin resonance spectroscopy under
  ambient conditions}.
\newblock \emph{\bibinfo{journal}{Science}} \textbf{\bibinfo{volume}{347}},
  \bibinfo{pages}{1135--1138} (\bibinfo{year}{2015}).

\bibitem{Schlipf2017}
\bibinfo{author}{Schlipf, L.} \emph{et~al.}
\newblock \bibinfo{title}{A molecular quantum spin network controlled by a
  single qubit}.
\newblock \emph{\bibinfo{journal}{Sci. Adv.}} \textbf{\bibinfo{volume}{3}},
  \bibinfo{pages}{e1701116} (\bibinfo{year}{2017}).

\bibitem{Shi2018}
\bibinfo{author}{Shi, F.} \emph{et~al.}
\newblock \bibinfo{title}{Single-{DNA} electron spin resonance spectroscopy in
  aqueous solutions}.
\newblock \emph{\bibinfo{journal}{Nat. Methods}} \textbf{\bibinfo{volume}{15}},
  \bibinfo{pages}{697--699} (\bibinfo{year}{2018}).

\bibitem{Pinto2020}
\bibinfo{author}{Pinto, D.} \emph{et~al.}
\newblock \bibinfo{title}{Readout and control of an endofullerene electronic
  spin}.
\newblock \emph{\bibinfo{journal}{Nat. Commun.}} \textbf{\bibinfo{volume}{11}},
  \bibinfo{pages}{6405} (\bibinfo{year}{2020}).

\bibitem{Barry2020}
\bibinfo{author}{Barry, J.~F.} \emph{et~al.}
\newblock \bibinfo{title}{Sensitivity optimization for nv-diamond
  magnetometry}.
\newblock \emph{\bibinfo{journal}{Rev. Mod. Phys.}}
  \textbf{\bibinfo{volume}{92}}, \bibinfo{pages}{015004}
  (\bibinfo{year}{2020}).

\bibitem{Wolf2015}
\bibinfo{author}{Wolf, T.} \emph{et~al.}
\newblock \bibinfo{title}{Subpicotesla diamond magnetometry}.
\newblock \emph{\bibinfo{journal}{Phys. Rev. X}} \textbf{\bibinfo{volume}{5}},
  \bibinfo{pages}{041001} (\bibinfo{year}{2015}).

\bibitem{Zhou2020}
\bibinfo{author}{Zhou, H.} \emph{et~al.}
\newblock \bibinfo{title}{Quantum metrology with strongly interacting spin
  systems}.
\newblock \emph{\bibinfo{journal}{Phys. Rev. X}} \textbf{\bibinfo{volume}{10}},
  \bibinfo{pages}{031003} (\bibinfo{year}{2020}).

\bibitem{Zhang2021}
\bibinfo{author}{Zhang, C.} \emph{et~al.}
\newblock \bibinfo{title}{Diamond magnetometry and gradiometry towards
  subpicotesla dc field measurement}.
\newblock \emph{\bibinfo{journal}{Phys. Rev. Applied}}
  \textbf{\bibinfo{volume}{15}}, \bibinfo{pages}{064075}
  (\bibinfo{year}{2021}).

\bibitem{Wang2022}
\bibinfo{author}{Wang, Z.} \emph{et~al.}
\newblock \bibinfo{title}{Picotesla magnetometry of microwave fields with
  diamond sensors}.
\newblock \emph{\bibinfo{journal}{Sci. Adv.}} \textbf{\bibinfo{volume}{8}},
  \bibinfo{pages}{eabq8158} (\bibinfo{year}{2022}).

\bibitem{Grinolds2013}
\bibinfo{author}{Grinolds, M.~S.} \emph{et~al.}
\newblock \bibinfo{title}{Nanoscale magnetic imaging of a single electron spin
  under ambient conditions}.
\newblock \emph{\bibinfo{journal}{Nat. Phys.}} \textbf{\bibinfo{volume}{9}},
  \bibinfo{pages}{215--219} (\bibinfo{year}{2013}).

\bibitem{Simpson2017}
\bibinfo{author}{Simpson, D.~A.} \emph{et~al.}
\newblock \bibinfo{title}{Electron paramagnetic resonance microscopy using
  spins in diamond under ambient conditions}.
\newblock \emph{\bibinfo{journal}{Nat. Commun.}} \textbf{\bibinfo{volume}{8}},
  \bibinfo{pages}{458} (\bibinfo{year}{2017}).

\bibitem{Grant2023}
\bibinfo{author}{Grant, E.~S.} \emph{et~al.}
\newblock \bibinfo{title}{Method for in-solution, high-throughput ${T}_{1}$
  relaxometry using fluorescent nanodiamonds}.
\newblock \emph{\bibinfo{journal}{Phys. Rev. Appl.}}
  \textbf{\bibinfo{volume}{20}}, \bibinfo{pages}{034018}
  (\bibinfo{year}{2023}).

\bibitem{iyer_2024}
\bibinfo{author}{Iyer, S.} \emph{et~al.}
\newblock \bibinfo{title}{Optically-trapped-nanodiamond relaxometric detection
  of nanomolar paramagnetic spins in aqueous environments}.
\newblock \emph{\bibinfo{journal}{Phys. Rev. Appl.}}
  \textbf{\bibinfo{volume}{22}}, \bibinfo{pages}{064076}
  (\bibinfo{year}{2024}).

\bibitem{Borbat2001}
\bibinfo{author}{Borbat, P.~P.}, \bibinfo{author}{Costa-Filho, A.~J.},
  \bibinfo{author}{Earle, K.~A.}, \bibinfo{author}{Moscicki, J.~K.} \&
  \bibinfo{author}{Freed, J.~H.}
\newblock \bibinfo{title}{Electron spin resonance in studies of membranes and
  proteins}.
\newblock \emph{\bibinfo{journal}{Science}} \textbf{\bibinfo{volume}{291}},
  \bibinfo{pages}{266--269} (\bibinfo{year}{2001}).

\bibitem{Kong2018}
\bibinfo{author}{Kong, F.} \emph{et~al.}
\newblock \bibinfo{title}{Nanoscale zero-field electron spin resonance
  spectroscopy}.
\newblock \emph{\bibinfo{journal}{Nat. Commun.}} \textbf{\bibinfo{volume}{9}},
  \bibinfo{pages}{1563} (\bibinfo{year}{2018}).

\bibitem{Qin2023}
\bibinfo{author}{Qin, Z.} \emph{et~al.}
\newblock \bibinfo{title}{In situ electron paramagnetic resonance spectroscopy
  using single nanodiamond sensors}.
\newblock \emph{\bibinfo{journal}{Nat. Commun.}} \textbf{\bibinfo{volume}{14}},
  \bibinfo{pages}{6278} (\bibinfo{year}{2023}).

\bibitem{Bogle1961}
\bibinfo{author}{Bogle, G.~S.}, \bibinfo{author}{Symmons, H.~F.},
  \bibinfo{author}{Burgess, V.~R.} \& \bibinfo{author}{Sierins, J.~V.}
\newblock \bibinfo{title}{Paramagnetic resonance spectrometry at zero magnetic
  field}.
\newblock \emph{\bibinfo{journal}{Proc. Phys. Soc.}}
  \textbf{\bibinfo{volume}{77}}, \bibinfo{pages}{561} (\bibinfo{year}{1961}).

\bibitem{Cole1963}
\bibinfo{author}{Cole, T.}, \bibinfo{author}{Kushida, T.} \&
  \bibinfo{author}{Heller, H.~C.}
\newblock \bibinfo{title}{Zero-field electron magnetic resonance in some
  inorganic and organic radicals}.
\newblock \emph{\bibinfo{journal}{J. Phys. Chem.}}
  \textbf{\bibinfo{volume}{38}}, \bibinfo{pages}{2915--2924}
  (\bibinfo{year}{1963}).

\bibitem{Bramley1983}
\bibinfo{author}{Bramley, R.} \& \bibinfo{author}{Strach, S.~J.}
\newblock \bibinfo{title}{Electron paramagnetic resonance spectroscopy at zero
  magnetic field}.
\newblock \emph{\bibinfo{journal}{Chem. Rev.}} \textbf{\bibinfo{volume}{83}},
  \bibinfo{pages}{49--82} (\bibinfo{year}{1983}).

\bibitem{SOM}
\emph{\bibinfo{title}{See Supplemental Material at
  http://link.aps.org/supplemental/XXX for detailed discussion on the theory,
  setup, samples, X-band EPR results, and baseline subtraction method.}}

\bibitem{kianinia_quantum_2022}
\bibinfo{author}{Kianinia, M.}, \bibinfo{author}{Xu, Z.-Q.},
  \bibinfo{author}{Toth, M.} \& \bibinfo{author}{Aharonovich, I.}
\newblock \bibinfo{title}{Quantum emitters in {2D} materials: {Emitter}
  engineering, photophysics, and integration in photonic nanostructures}.
\newblock \emph{\bibinfo{journal}{Appl. Phys. Rev.}}
  \textbf{\bibinfo{volume}{9}}, \bibinfo{pages}{011306} (\bibinfo{year}{2022}).

\bibitem{Xu2019}
\bibinfo{author}{Xu, L.} \emph{et~al.}
\newblock \bibinfo{title}{High-efficiency fluorescence collection for
  {NV}$^{-}$ center ensembles in diamond}.
\newblock \emph{\bibinfo{journal}{Opt. Express}} \textbf{\bibinfo{volume}{27}},
  \bibinfo{pages}{10787--10797} (\bibinfo{year}{2019}).

\end{thebibliography}

\bibliographystyle{naturemag}

\end{document}


\title{Parallel accelerated electron paramagnetic resonance spectroscopy using diamond sensors}

\author{Zhehua Huang}
\altaffiliation{These authors contributed equally to this work.}
\affiliation{CAS Key Laboratory of Microscale Magnetic Resonance and School of Physical Sciences, University of Science and Technology of China, Hefei 230026, China}
\affiliation{CAS Center for Excellence in Quantum Information and Quantum Physics, University of Science and Technology of China, Hefei 230026, China}

\author{Zhengze Zhao}
\altaffiliation{These authors contributed equally to this work.}
\affiliation{CAS Key Laboratory of Microscale Magnetic Resonance and School of Physical Sciences, University of Science and Technology of China, Hefei 230026, China}
\affiliation{CAS Center for Excellence in Quantum Information and Quantum Physics, University of Science and Technology of China, Hefei 230026, China}

\author{Fei Kong}
\email{kongfei@ustc.edu.cn}
\affiliation{CAS Key Laboratory of Microscale Magnetic Resonance and School of Physical Sciences, University of Science and Technology of China, Hefei 230026, China}
\affiliation{CAS Center for Excellence in Quantum Information and Quantum Physics, University of Science and Technology of China, Hefei 230026, China}
\affiliation{Hefei National Laboratory, University of Science and Technology of China, Hefei 230088, China}

\author{Zhecheng Wang}
\affiliation{CAS Key Laboratory of Microscale Magnetic Resonance and School of Physical Sciences, University of Science and Technology of China, Hefei 230026, China}
\affiliation{CAS Center for Excellence in Quantum Information and Quantum Physics, University of Science and Technology of China, Hefei 230026, China}
\affiliation{School of Biomedical Engineering and Suzhou Institute for Advanced Research, University of Science and Technology of China, Suzhou 215123, China}

\author{Pengju Zhao}
\affiliation{CAS Key Laboratory of Microscale Magnetic Resonance and School of Physical Sciences, University of Science and Technology of China, Hefei 230026, China}
\affiliation{CAS Center for Excellence in Quantum Information and Quantum Physics, University of Science and Technology of China, Hefei 230026, China}

\author{Xiangtian Gong}
\affiliation{CAS Key Laboratory of Microscale Magnetic Resonance and School of Physical Sciences, University of Science and Technology of China, Hefei 230026, China}
\affiliation{CAS Center for Excellence in Quantum Information and Quantum Physics, University of Science and Technology of China, Hefei 230026, China}

\author{Xiangyu Ye}
\affiliation{CAS Key Laboratory of Microscale Magnetic Resonance and School of Physical Sciences, University of Science and Technology of China, Hefei 230026, China}
\affiliation{CAS Center for Excellence in Quantum Information and Quantum Physics, University of Science and Technology of China, Hefei 230026, China}

\author{Ya Wang}
\affiliation{CAS Key Laboratory of Microscale Magnetic Resonance and School of Physical Sciences, University of Science and Technology of China, Hefei 230026, China}
\affiliation{CAS Center for Excellence in Quantum Information and Quantum Physics, University of Science and Technology of China, Hefei 230026, China}
\affiliation{Hefei National Laboratory, University of Science and Technology of China, Hefei 230088, China}

\author{Fazhan Shi}
\email{fzshi@ustc.edu.cn}
\affiliation{CAS Key Laboratory of Microscale Magnetic Resonance and School of Physical Sciences, University of Science and Technology of China, Hefei 230026, China}
\affiliation{CAS Center for Excellence in Quantum Information and Quantum Physics, University of Science and Technology of China, Hefei 230026, China}
\affiliation{Hefei National Laboratory, University of Science and Technology of China, Hefei 230088, China}
\affiliation{School of Biomedical Engineering and Suzhou Institute for Advanced Research, University of Science and Technology of China, Suzhou 215123, China}

\author{Jiangfeng Du}
\email{djf@ustc.edu.cn}
\affiliation{CAS Key Laboratory of Microscale Magnetic Resonance and School of Physical Sciences, University of Science and Technology of China, Hefei 230026, China}
\affiliation{CAS Center for Excellence in Quantum Information and Quantum Physics, University of Science and Technology of China, Hefei 230026, China}
\affiliation{Hefei National Laboratory, University of Science and Technology of China, Hefei 230088, China}
\affiliation{School of Physics, Zhejiang University, Hangzhou 310027, China}

\appendix
\onecolumngrid
\clearpage
\section{Supplemental materials}
\subsection{Supplementary Note 1. The theory of the zero-field EPR spectroscopy}
As described in the main text, we obtain the spectral line information under zero bias field by applying an amplitude-modulated microwave driving, fixing the evolution time, and sweeping the modulation frequency $f$.

This sequence had been described in detail in our previous article (see details in \cite{Qin2023}), where the Hamiltonian of the NV and the target spin system is considered to be,

\begin{equation}\label{eq:Hamiltonian2leveltarget}
\Hamil = D_\mathrm{gs} S_z^2 + \gamma_\mathrm{NV} B_1 \sin\theta \cos ft \cos D_\mathrm{gs} t \,S_x + \sum_{i,j=\qty{x,y,z}} \mathcal{D}_{ij} S_i T_j + \omega T_z.
\end{equation}
Here $\bm{S}$ is the spin operator for the NV electron spin, and $\bm{T}$ is the spin operator of the effective two-level-system in the hyperfine-coupled target spin Hamiltonian. (Notations: $\gamma_\mathrm{NV} = \SI{-28.03}{\GHz/\tesla}$, $D_\mathrm{gs} = \SI{2.87}{\GHz}$.)

After several steps of unitary rotation and simplification, the Hamiltonian can be written as,
\begin{equation}
\Hamil_{\mathrm{III}} \approx \mathcal{D}_{zz} S_z T_z + (\omega - f) T_z - \frac{\kappa}{4}\mathcal{D}_{zx}S_y T_y + \frac{\kappa}{4}\mathcal{D}_{zy}S_y T_x,
\end{equation}
where $\kappa = \Omega/f = \gamma_\mathrm{NV} B_1\sin\theta /f$.

At the resonance condition $f = \omega$, the NV center will have an additional longitudinal relaxation (denoted by $\Delta\Gamma_1$). The population of state $\ket{0}$ after a evolution time $t$ is
\begin{equation}
P_0(f,t) = \frac{1}{3} + \frac{2}{3}\exp[-(\Gamma_1^\prime + \Delta\Gamma_1)t]\qcomma 
\Delta\Gamma_1 = \frac{3\kappa^2(\mathcal{D}_{zx}^2+\mathcal{D}_{zy}^2)}{64}\frac{\Gamma_2}{\Gamma_2^2+(f-\omega)^2},
\end{equation}
where $\Gamma_2 = \Gamma_{2,\textrm{NV}} + \Gamma_{2,\textrm{target}}$. (Notations: $\Gamma_1^\prime = 1/T_{1,\mathrm{NV}}^\prime$ is the  longitudinal relaxation rate of the NV center under amplitude-modulation driving. $\Gamma_{2,\textrm{NV}}$ and $\Gamma_{2,\textrm{target}}$ are the spin decoherence rate of NV center and the target spin, respectively.)

\revisen{The observed signal is the population difference between the presence and absence of $\Delta\Gamma_1$. Since $\Delta\Gamma_1 t \ll 1$ is typically a small quantity, the signal can be approximated as $S(f,t) \approx \frac{2}{3}(\Delta\Gamma_1 t)\exp(-\Gamma_1^\prime t)$.  Setting $\dv{S(f,t)}{t} = 0$ gives the maximum value at $t = 1/\Gamma_1^\prime$.}

\begin{figure}[htbp]
\centering \includegraphics[width=1\textwidth]{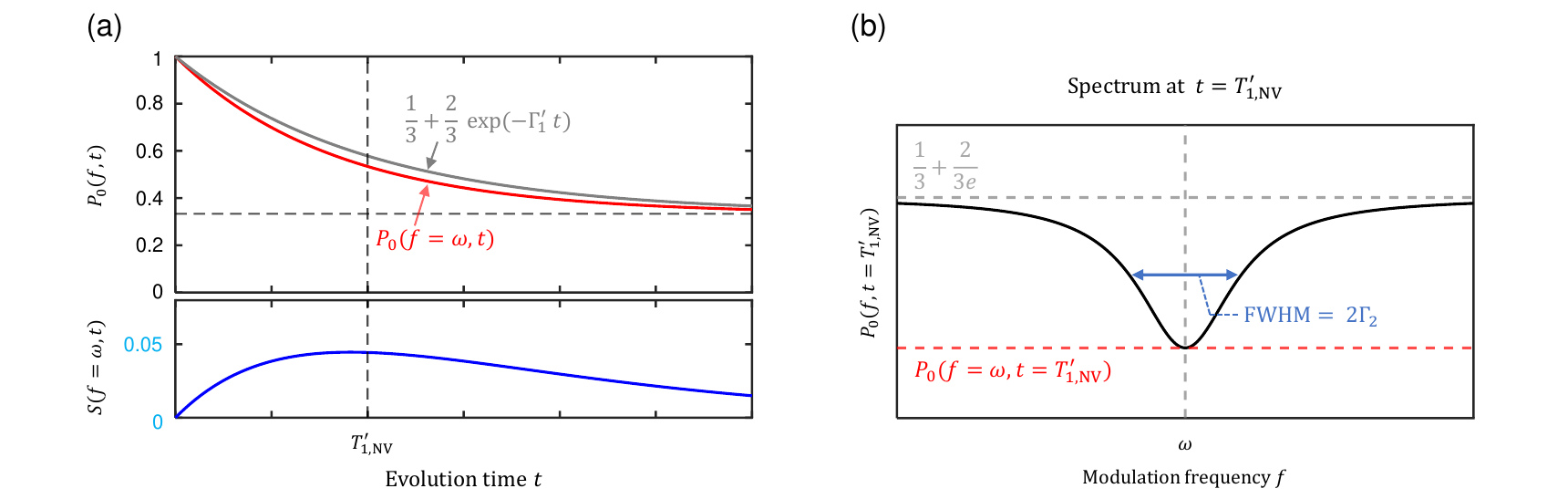}
\protect\caption{\textbf{Dependence of signal contrast on time and modulation frequency.}
	\revisen{(a) (Upper panel) The red/gray line depicts the theoretical time evolution of the $\ket{0}$-state population when the resonance condition is met/not met, respectively. (Lower Panel) The blue line represents the difference between these two cases (i.e., the signal contrast), which peaks at $t=T_{1,\textrm{NV}}^\prime$.
	(b) The theoretical spectrum readout at time $t=T_{1,\textrm{NV}}^\prime$. Its linewidth is $2\Gamma_2$,  equivalent to the sum of the NV and target linewidths.}
	}
	\label{fig:SignalContrast}
\end{figure}

Finally, the maximum signal contrast at optimum evolution time $t = T_{1,\mathrm{NV}}^\prime$ is,
\begin{equation}\label{eq:EnsSignalContrast}
S(f = \omega,t = T_{1,\mathrm{NV}}^\prime) \approx \frac{\kappa^2(\mathcal{D}_{zx}^2+\mathcal{D}_{zy}^2)}{32}\frac{1}{\ee\Gamma_1^\prime\Gamma_2}.
\end{equation}

\clearpage
\subsection{Supplementary Note 2. The ensemble averaged signals of nitroxide radicals}
Next, we consider the relative coupling strength of different 2-level transitions that are selection-rule-allowed for the nitroxide electron spins. Here we follow the same steps and notations as the article\cite{Qin2023} before, and generalize the calculation to include the $^{14}$N nitroxide radicals, where the nuclear spin number is an integer.

In the NV axis frame $\qty{\hat{x},\hat{y},\hat{z}}$, define the target spin position vector relative to NV as $\vec{r} = r\cdot\hat{r}(\theta_r,\phi_r)$, and define the unit direction vector parallel to the principle axis of target spin as $\hat{n}_\mathrm{e} = \hat{n}_\mathrm{e}(\theta_\mathrm{e},\phi_\mathrm{e})$. The full Hamiltonian of the sensor-target system is,
\begin{equation}
\begin{aligned}
\Hamil =& \Hamil_\mathrm{NV} + \Hamil_\mathrm{int} + \Hamil_\mathrm{T}\\
=& D S_z^2 + \gamma_\mathrm{NV} B_1 \sin\theta \cos ft \cos Dt \,S_x + \vec{S}\mathcal{D}_0 \mathcal{R}\vec{S}^\prime
+ A_\perp(S_x^\prime I_x + S_y^\prime I_y) + A_\para S_z^\prime I_z + P[I_z^2-\frac{1}{3}I(I+1)].
\end{aligned}
\end{equation}
The nuclear quadrupole interaction constant $P$ is non-zero for the $^{14}\textrm{N}$ nuclear, but its value is small ($\leq \SI{2}{\MHz}$) compared to the hyperfine constant ($20\sim\SI{100}{\MHz}$). Therefore, we ignore the quadrupole term in the following discussion.

The spin Hamiltonian of the target spin system $\Hamil_\mathrm{T}$ gives the following eigen-states,
\begin{equation}
\ket{T,m_T} = \cos{\frac{\alpha_{m_T}}{2}}\ket{\frac{1}{2},m_T-\frac{1}{2}}+\sin{\frac{\alpha_{m_T}}{2}}\ket{-\frac{1}{2},m_T+\frac{1}{2}}\qcomma m_T = -I+1/2,-I+3/2,\cdots,I-1/2,
\end{equation}
where $\alpha_{m_T}$ should satisfy the condition that, 

\begin{equation}
\tan\alpha_{m_T} = \frac{A_\perp}{A_\para}\cdot\frac{\sqrt{I(I+1)-(m_T-1/2)(m_T+1/2)}}{m_T} = \frac{\chi_{m_T}}{m_T}.
\end{equation}

For a single $m_T$, there are two different eigenstates (one $\tan\alpha$ value corresponds to two different $\tan\alpha/2$ values), and their corresponding energy levels are
\begin{equation}\label{eq:HyperfineEnergyLevels}
E_{\uparrow,m_T} = \frac{A_\para}{2}\qty[\sqrt{\chi_{m_T}^2+m_T^2}-\frac{1}{2}],
\textrm{and }
E_{\downarrow,m_T} = \frac{A_\para}{2}\qty[-\sqrt{\chi_{m_T}^2+m_T^2}-\frac{1}{2}].
\end{equation}

We here call the energy levels with energy higher than $-\frac{A_\para}{4}$ the `upper branch', marked by $\uparrow$, and the group with energy lower than $-\frac{A_\para}{4}$ the `lower branch', marked by $\downarrow$. (Eigen-states $\ket{\uparrow,\frac{1}{2}+I} = \ket{\frac{1}{2},I}$ and $\ket{\uparrow,-\frac{1}{2}-I} = \ket{-\frac{1}{2},-I}$ are two special cases that has only upper branch levels. However, they are still well described by the formula\eqref{eq:HyperfineEnergyLevels}, when setting $m_T = \pm(\frac{1}{2}+I)$). 

By re-expressing the target spin operator in the basis of its hyperfine Hamiltonian eigenstates, and choosing only the two eigenstates that participates in the interested transition, the Hamiltonian is effectively written as \eqref{eq:Hamiltonian2leveltarget}, while,
\begin{equation}
\begin{gathered}
\mathcal{D} = \mathcal{D}_0\cdot \mathcal{R}\cdot \mathcal{M}\qcomma\\
\mathcal{D}_0 = \frac{\mu_0\gamma_\mathrm{NV}\gamma_e\hbar}{4\pi r^3}(\mathbf{1}_{3\times3}-3\hat{\bm{r}}^\mathsf{T}\hat{\bm{r}}),\\
\mathcal{R}(\hat{n}(\theta_\mathrm{e},\phi_\mathrm{e})) = 
\begin{pmatrix}
\cos\theta_\mathrm{e}\cos\phi_\mathrm{e}&-\sin\phi_\mathrm{e}&\sin\theta_\mathrm{e}\cos\phi_\mathrm{e}\\
\cos\theta_\mathrm{e}\sin\phi_\mathrm{e}&\cos\phi_\mathrm{e}&\sin\theta_\mathrm{e}\sin\phi_\mathrm{e}\\
-\sin\theta_\mathrm{e}&0&\cos\theta_\mathrm{e}
\end{pmatrix}\qcomma\\
\mathcal{M}_0 = \cos\frac{\alpha_0+\alpha_1}{2}\mqty(0&0&0\\0&0&0\\1&0&\ast)\qcomma
\mathcal{M}_1 = \sin\frac{\alpha_0}{2}\cos\frac{\alpha_1}{2}\mqty(1&0&0\\0&1&0\\0&0&\ast)\qcomma
\mathcal{M}_{-1} = \cos\frac{\alpha_0}{2}\sin\frac{\alpha_1}{2}\mqty(1&0&0\\0&-1&0\\0&0&\ast)\qcomma
\end{gathered}
\end{equation}
where the $\alpha_0,\alpha_1$ are the $\alpha_{m_T}$ value of a specific transition $\ket{T_0,m_{T_0}}\leftrightarrow\ket{T_1,m_{T_1}}$.

The coefficient of the reduction matrix $\mathcal{M}$ clearly shows that different resonance transitions may have different relative strengths, which is independent on the target position and orientation. These relative strengths are quantified by $\xi(m_T)$, defined as,
\begin{equation}
\xi(m_T) = 
    \begin{cases}
      \cos^2\frac{\alpha_{m_T}+\alpha_{m_{T}+1}}{2}, & \Delta m_T = 0\\
      \sin^2\frac{\alpha_{m_T}}{2}\cos^2\frac{\alpha_{m_{T}+1}}{2}, & \Delta m_T = 1\\
      \cos^2\frac{\alpha_{m_T}}{2}\sin^2\frac{\alpha_{m_{T}+1}}{2}, & \Delta m_T = -1\\
    \end{cases}.       
\end{equation}
Similar to \cite{Qin2023}, we can sum up the $\xi(m_T)$ that shares the same resonance frequency. This can finally be summarized to the following table,
\begin{equation}
\arraycolsep=5pt\def\arraystretch{2.5}
\begin{array}{r|c|c}
		\hline\hline
		\xi(m_T) & I\text{: half Integer} & I\text{: Integer}\\
		\hline
		\Delta m_T = 0 & \displaystyle\frac{\tan^2\alpha_{m_T}}{1+\tan^2\alpha_{m_T}}\times
		\begin{cases}
		1,&m_T = 0\\
		2,&m_T \neq 0
		\end{cases}
		& \displaystyle\frac{\tan^2\alpha_{m_T}}{1+\tan^2\alpha_{m_T}}\\\hline
		\Delta m_T = \pm 1\qcomma\text{in branch} &
		1-\cos\alpha^\uparrow_\abs{m_T}\cos\alpha^\uparrow_\abs{m_T+1} &
		\begin{cases}
		(1+\cos\alpha^\uparrow_\abs{m_T}\cos\alpha^\uparrow_\abs{m_T+1})/2&m_T = -1/2\\
		1-\cos\alpha^\uparrow_\abs{m_T}\cos\alpha^\uparrow_\abs{m_T+1} & m_T \neq -1/2\\
		\end{cases}\\\hline
		\Delta m_T = \pm 1\qcomma\text{between branch} &
		1+\cos\alpha^\uparrow_\abs{m_T}\cos\alpha^\uparrow_\abs{m_T+1} & 
		\begin{cases}
		(1-\cos\alpha^\uparrow_\abs{m_T}\cos\alpha^\uparrow_\abs{m_T+1})/2&m_T = -1/2\\
		1+\cos\alpha^\uparrow_\abs{m_T}\cos\alpha^\uparrow_\abs{m_T+1} & m_T \neq -1/2\\
		\end{cases}\\\hline\hline
	\end{array}
\end{equation}
(Note: The derivation can be performed using the trigonometric identity that $\alpha_{m_T}$ satisfies, $\cos\alpha_{m_T}^\uparrow = \cos\alpha_{-m_T}^\uparrow = \cos\alpha_\abs{m_T}^\uparrow$, and $\cos\alpha_{m_T}^\downarrow = - \cos\alpha_{m_T}^\uparrow$).

We assume that the NV center lies under the bulk diamond surface with a depth $h$, the resonant $\Delta\Gamma_1$(at $f = \omega$) averaged over all positions and orientations of target molecule homogeneously distributed in the half space above diamond surface is,
\begin{equation}
\expval{\Delta\Gamma_1} = 
\int_h^\infty c\cdot r^2\dd r \int_0^{\arccos\frac{h}{r}}\sin\theta\dd\theta \int_0^{2\pi}\dd\phi
\int_0^\pi\frac{\sin\theta_e\dd\theta_e}{2} \int_0^{2\pi}\frac{\dd\phi_e}{2\pi}
\frac{1}{2I+1}\frac{3\kappa^2((Q\mathcal{D})_{zx}^2+(Q\mathcal{D})_{zy}^2)}{64\Gamma_2},
\end{equation}
where 
\begin{equation}
Q = \mqty(\cos\alpha&0&-\sin\alpha\\0&1&0\\\sin\alpha&0&\cos\alpha)
\end{equation}
is a rotation from the NV reference frame to the laboratory frame, and $\alpha$ is the angle between the NV axis ($\alpha = \acos(\pm 1/\sqrt{3})$ for a $\qty{100}$-surface) and the diamond surface normal. $c$ is the concentration of the target molecules.
The integral gives,
\begin{equation}
\expval{\Delta\Gamma_1}
= \frac{1}{2I+1}\qty(\frac{\mu_0\gamma_\mathrm{NV}\gamma_e\hbar}{4\pi})^2\times\frac{\pi\beta\xi c\kappa^2}{256\,\Gamma_2 h^3}(3+\cos2\alpha).
\end{equation}

Finally, we come to see the ensemble overall-averaged signal contrast, recalling equation\eqref{eq:EnsSignalContrast},
\begin{equation}\label{eq:EnsembleOverallContrast}
S(f=\omega,t=T_{1,\mathrm{NV}}^\prime) \approx \frac{1}{2I+1}\qty(\frac{\mu_0\gamma_\mathrm{NV}\gamma_e\hbar}{4\pi})^2\times\frac{\pi\beta\xi c\kappa^2}{384\, h^3}(3+\cos2\alpha)\times\frac{1}{\ee\Gamma_1^\prime\Gamma_2}.
\end{equation}

\begin{figure}[htbp]
\centering \includegraphics[width=1\textwidth]{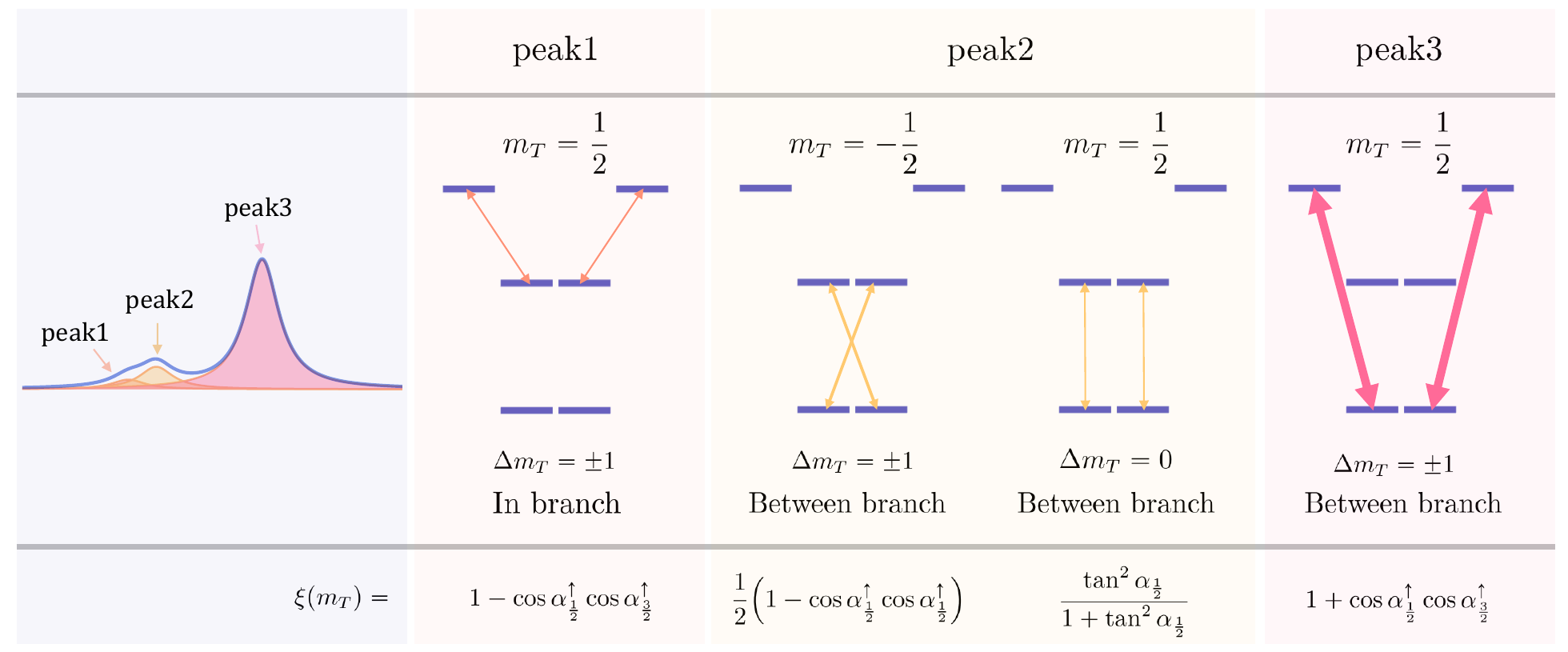}
\protect\caption{\textbf{Theoretical zero-field EPR spectrum and corresponding transitions for $^{14}$N nitroxide radicals.}
	The thickness of the arrow qualitatively indicates the strength of the transition, with peak3 significantly stronger than the other two peaks.
	}
	\label{fig:N14EnergyLevelsAndTransitions}
\end{figure}
\begin{figure}[htbp]
\centering \includegraphics[width=0.8\textwidth]{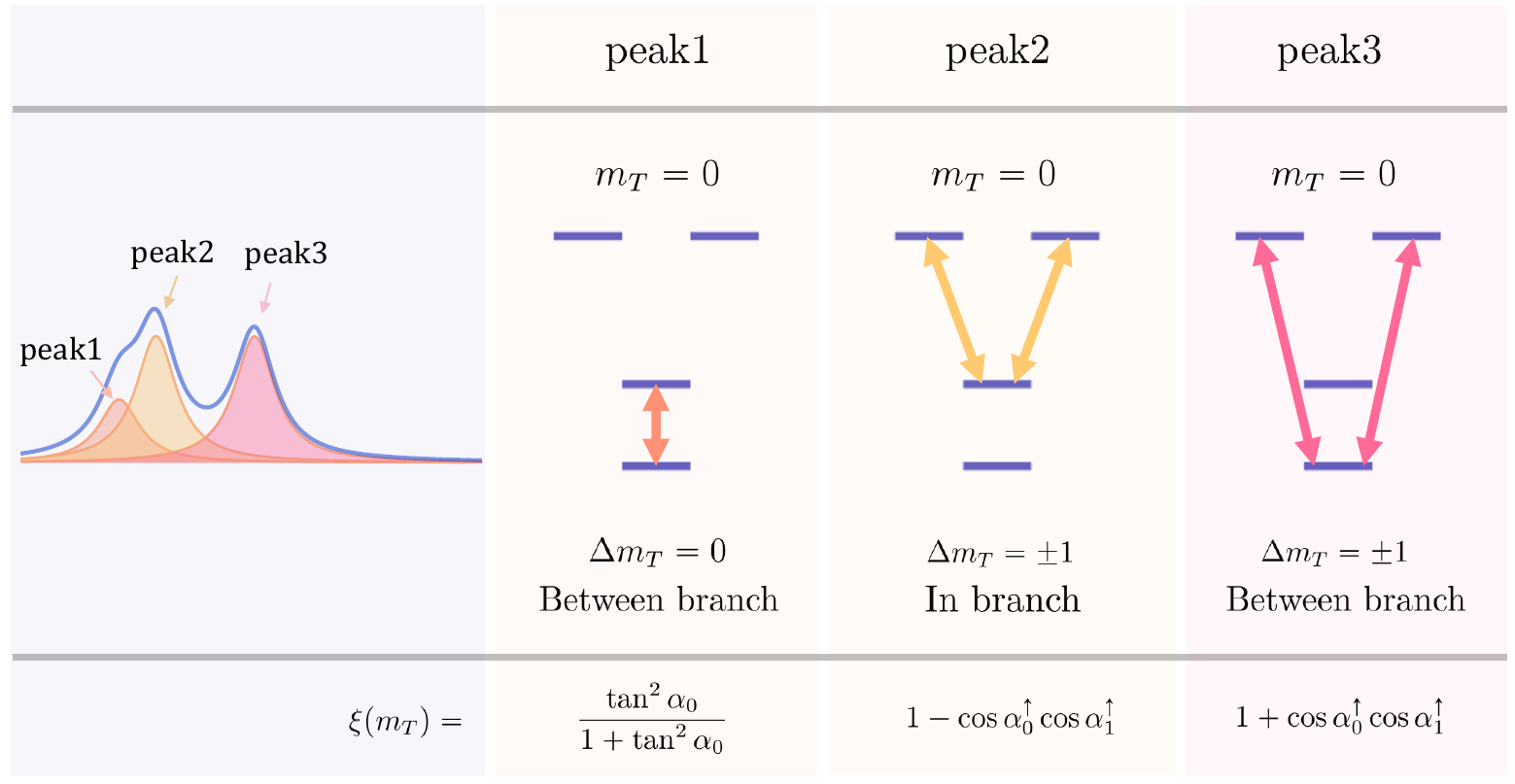}
\protect\caption{\textbf{Theoretical zero-field EPR spectrum and corresponding transitions for $^{15}$N nitroxide radicals.}
	The thickness of the arrow qualitatively indicates the strength of the transition. For the $^{15}$N nitroxide, the height of each peak is approximately $1:2:2$.
	}
	\label{fig:N15EnergyLevelsAndTransitions}
\end{figure}

With the hyperfine constants obtained from the conventional X-band EPR measurement (see Note 4), we can calculate the zero-field EPR spectrum according to formula\eqref{eq:EnsembleOverallContrast} (see Fig.~\ref{fig:N14EnergyLevelsAndTransitions}, Fig.~\ref{fig:N15EnergyLevelsAndTransitions}). We mark the transitions corresponding to each peak with arrows in the corresponding energy level diagram. 
\revisen{The solution to the spin Hamiltonian of the target spin system $\Hamil_\mathrm{T}$ yields the following transition spectrum peaks: $\omega_1=\frac{3}{4}A_\para-\frac{1}{4}\sqrt{8A_\perp^2 + A_\para^2}$, $\omega_2 = \frac{1}{2}\sqrt{8A_\perp^2 + A_\para^2}$, $\omega_3 =\frac{3}{4}A_\para+\frac{1}{4}\sqrt{8A_\perp^2 + A_\para^2}$ for $^{14}\mathrm{N}$ TEMPO, and $\omega_1 = A_\perp$, $\omega_2 = \frac{1}{2}(A_\para-A_\perp)$, $\omega_3 = \frac{1}{2}(A_\para+A_\perp)$ for $^{15}\mathrm{N}$ TEMPO.}
The explicit expression of the relative strength coefficient $\xi$ of the corresponding transition are listed at the bottom.
\clearpage

\subsection{Supplementary Note 3. The EPR system}
	\subsubsection{Experimental setup and diamond samples}
\begin{figure}[h]
		\centering
		\includegraphics[width=0.9\textwidth]{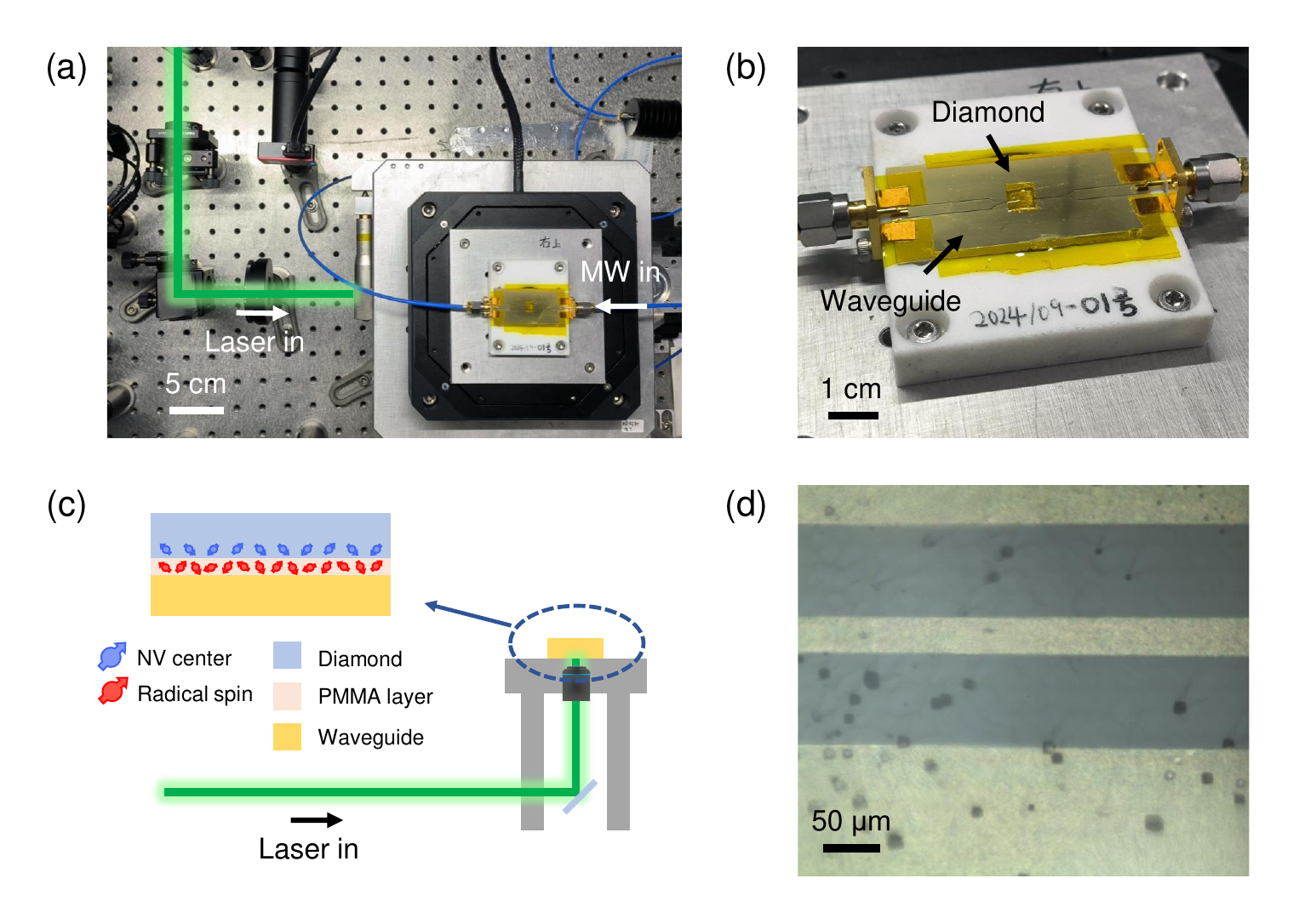}
		\caption{%
			\textbf{The EPR system.}
			\revisen{
			(a) Overall view of the EPR system. 
			(b) Optical image of the sample stage, including the diamond sample and a coplanar waveguide.
			(c) Structural diagram of the sample stage. The $\SI{532}{\nm}$ laser illuminates the sample through an objective positioned below the sample stage. The diamond containing NV centers, coated with radical samples on its surface, is mounted upside down on the waveguide. 
			(d) Microscope image of the diamond on the waveguide. 
			}
		}
		\label{fig:EPRsystem}
	\end{figure}

	\revisen{
	In the optical part (see Fig.~2(a) in the main text and Fig.~\ref{fig:EPRsystem}(a)), the excitation light source is a \SI{150}{\milli\watt} 532nm laser (CNI MGL-III-532-150mW).
	We place a convex lens with a focal length of \SI{70}{\milli\m} in front of the microscope objective (Olympus LUCPlanFL N 60x), so that the laser from the objective lens to the sample surface is a wide-field collimated light. The laser spot size is calibrated under the CCD (Thorlabs DCC1545M-GL, not shown here) as a Gaussian spot with a diameter of $d\sim\SI{17}{\micro\meter}$. The fluorescence of NV is collected by an avalanche photodiode (Excelitas Technologies SPCM-AQRH-14) after passing through the objective, dichroic mirror (Semrock Di03-R635-t3-25x36), and the optical filter (Semrock BLP01-635R-25).}
	
	\revisen{In the microwave part, the microwave sequence is generated by an arbitrary waveform generator (Keysight, M8190a), amplified by an microwave amplifier (Mini-Circuits, ZHL-16W-43-S+), and radiated to the NV center by a coplanar waveguide (see Fig.~\ref{fig:EPRsystem}(b)).}
	
	\revisen{The diamond sample, coated with radical samples on its surface, is mounted upside down on the coplanar waveguide (see Fig.~\ref{fig:EPRsystem}(b,c,d)). This configuration minimizes the distance between the NV centers/radical samples and the waveguide, thereby ensuring a sufficiently strong radiation field for spin manipulation. The coplanar waveguide is mounted on	a 3-axis positioning stage to facilitate the selection of the region required for the quenching experiment. }
	
	\revisen{The diamond sample we used is a 100-oriented sample with a $99.99\%$ $^{12}\textrm{C}$ isotopic purified surface layer. It is then implanted with $^{15}\mathrm{N}^+$ ions with an energy of \SI{5}{\keV}, dose $2\times10^{12}\SI{}{\per\cm\squared}$.}

\subsubsection{Radicals samples preparation}
\revisen{
All samples used in this article are obtained commercially, including nitroxyl radicals ($^{14}$N-TEMPO methacrylate, and $\text{4-Oxo-TEMPO-d}_{16}, ^{15}$N, free radical, Sigma-Aldrich) and poly(methyl methacrylate) (PMMA, mw=350,000, Sigma-Aldrich). The PMMA is dissolved in methylbenzene at a volume fraction of 1\%, while the nitroxides are dissolved in acetone at a concentration of \SI{100}{\milli\Molar}. The two solutions are then mixed in a ratio that achieved a final concentration of \SI{100}{\milli\Molar} nitroxides in solid-phase PMMA. The mixture is spin-coated onto the diamond surface at a speed of 4800 rpm to obtain a uniform layer. Prior to spin-coating, the diamond surface is cleaned by submerging it in a boiling mixture of nitric, perchloric, and sulfuric acids (in a 1:1:1 ratio) at 180°C for 4 hours.}

\subsection{Supplementary Note 4. X-band EPR measurement of nitroxide radicals}
To confirm that the radicals are uniformly dispersed in PMMA with restricted motion, we perform the X-band EPR measurement. The sample is similar with that for NV-EPR measurement. The only difference is the drying process: spin coating for NV-EPR and evaporation for X-band EPR. 

Specifically, the mixed solution is placed on a watch glass in a fume hood, allowing the solvent to evaporate. After the solvent has completely evaporated, a PMMA film is obtained, with radicals dispersed at a concentration of 100 mM. 

\begin{figure}[htbp]
\centering \includegraphics[width=0.95\textwidth]{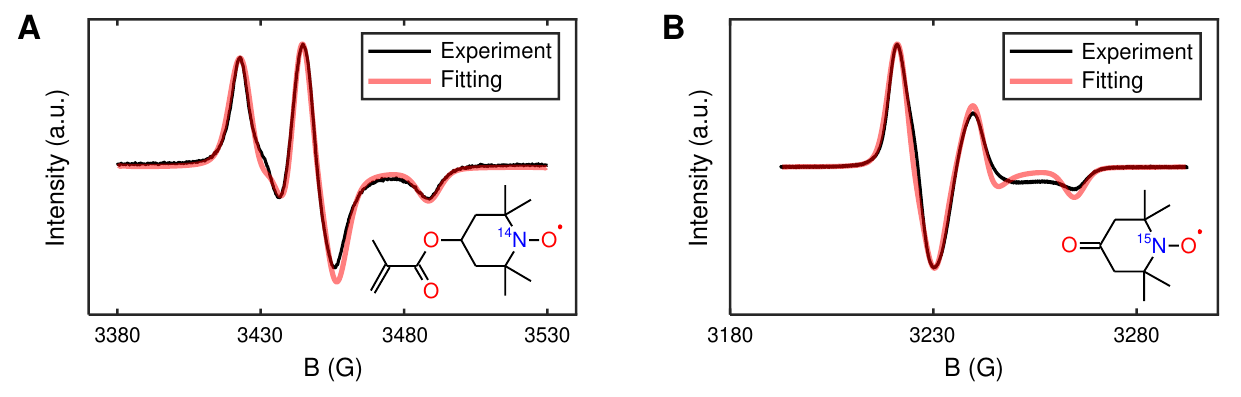}
\protect\caption{\textbf{X-band CW EPR of TEMPO dispersed in PMMA.}
	(\textbf{A}) EPR spectrum of $^{14}$N-TEMPO methacrylate (structure shown in inset, \SI{100}{\milli\Molar}), measured on a Bruker EMX spectrometer. The power and frequency of incident microwave is \SI{1}{\milli\watt}, \SI{9.67}{\GHz} respectively, field modulation \SI{1}{\gauss}, \SI{100}{\kHz}. The fitting result gives $A_\perp = \SI{19}{\MHz}$ and $A_\para = \SI{91}{\MHz}$. 
	(\textbf{B}) EPR spectrum of $\text{4-Oxo-TEMPO-d}_{16}, ^{15}$N (structure shown in inset, \SI{10}{\milli\Molar}), measured on a JES-FA200 spectrometer. The power and frequency of incident microwave is \SI{1}{\milli\watt}, \SI{9.07}{\GHz} respectively, field modulation \SI{0.2}{\gauss}, \SI{100}{\kHz}. The fitting result gives $A_\perp = \SI{32}{\MHz}$ and $A_\para = \SI{120}{\MHz}$. Both spectra are measured at room temperature, and are shown in the conventional derivative representation.}
	\label{fig:conventionalEPR}
\end{figure}

This film is then placed in the ensemble X-band EPR spectrometer to acquire the spectra of nitroxide radicals. These spectra are fitted by the EPR MATLAB software package
\textit{Easyspin}. As shown in Fig.~\ref{fig:conventionalEPR}, the measured spectrum is very similar to the powder spectrum, indicating that the motion of radicals is restricted. 

\clearpage
\section{Supplementary Note 5. Calibration of spectral baselines}
\begin{figure}[htbp]
	\centering
	\includegraphics[width=1\textwidth]{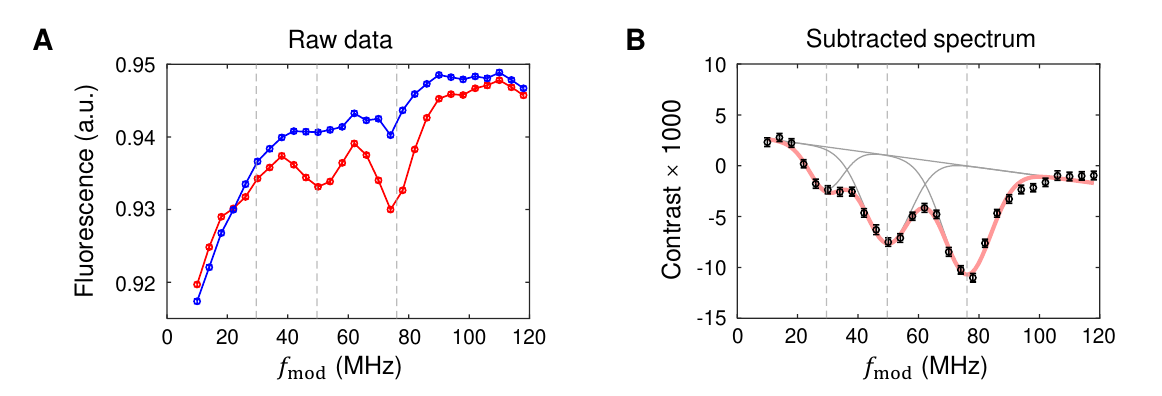}
	\caption{%
		(\textbf{A}) The raw data of the amplitude modulation spectra. The red(blue) dots are obtained before(after) the nitroxide radicals are quenched. Both spectra are measured under $\kappa = 0.7, \tau = 1$ ms and normalized by the counts in the $|0\rangle$ state.
		(\textbf{B}) The spectrum after background baseline subtraction. The black dots represent the signal of the $^{15}\text{N}$ nitroxide radicals, fitted by three Gaussian peaks on a slanted baseline, as shown by the light red line.
	}
	\label{fig:BackgroundSubtraction1}
\end{figure}

The red dots in Fig.~\ref{fig:BackgroundSubtraction1}A show the raw data of the amplitude modulation spectrum of $^{15}\text{N}$ nitroxide radicals. There exists a non-flat baseline with the unevenness reaching up to 3\% of the fluorescence, which hinders the measurement of spectrum. The non-flat baseline originates from the non-uniform frequency response of the waveguide to the microwave. After quenching the radicals using a high power density laser, the spectrum is represented by the blue dots, which reflects only the background baseline. 

By subtracting the data in blue dots from the data in red dots, three distinct peaks can be seen, as shown by the black dots in Fig.~\ref{fig:BackgroundSubtraction1}B. Three Gaussian peaks on a slanted baseline are used to fit the spectrum. These three peaks---centered at 30(2) MHz, 50(1) MHz, and 76(1) MHz, as illustrated by the vertical dashed lines---are much clearer in the subtracted spectrum than in the raw spectrum. The remaining slanted baseline is likely caused by the drift of the microwave power. In the main text Fig. 2, the slanted baseline is removed from the spectra.

\begin{figure}[htbp]
	\centering
	\includegraphics[width=1\textwidth]{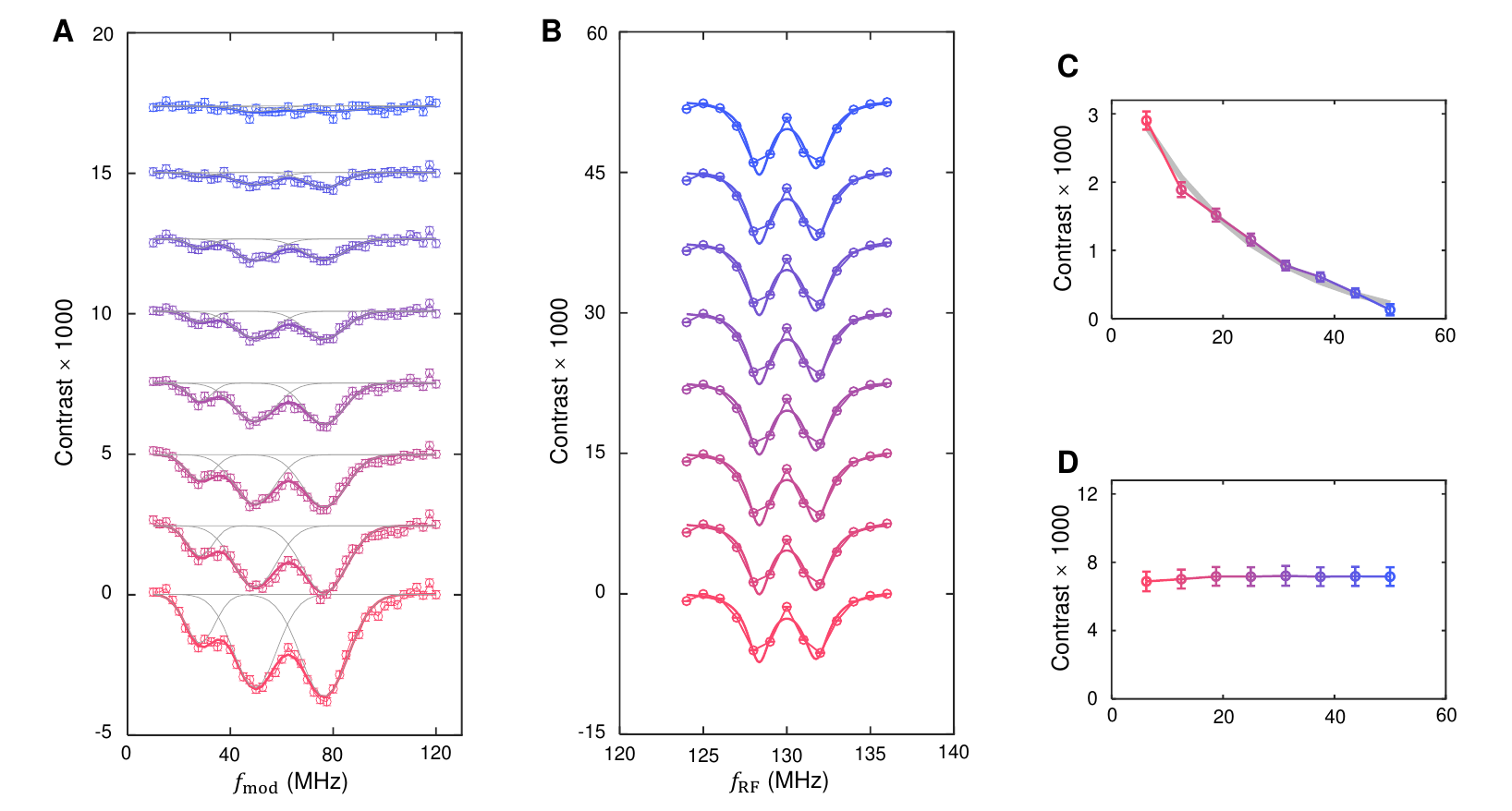}
	\caption{%
		(\textbf{A}) Real-time monitoring of the amplitude modulation spectra of $^{15}\text{N}$ nitroxide radicals, revealing a decreasing contrast. The spectra are obtained by sweeping $f_{\text{mod}}$, with $\kappa = 0.7$ and $\tau = 1$ ms.
		(\textbf{B}) The mean contrast of the three peaks from the nitroxide radicals during 50 hours of illumination with a $ 37 \text{ W}/\text{cm}^2 $ laser, showing an exponential decay with a characteristic time of 22(4) hours.
		(\textbf{C}) Simultaneous real-time monitoring of the amplitude modulation spectra of an artificial RF signal, exhibiting a stable contrast. The spectra are obtained by sweeping $f_{\text{RF}}$, with $f_{\text{mod}} = 130$ MHz, $\kappa = 0.42$ and $ \tau = 1 $ ms. 
		(\textbf{D}) The mean contrast of the two peaks from the artificial RF signal, showing a stable contrast during 50 hours of illumination with a $ 37 \text{ W}/\text{cm}^2 $ laser. 
	}
	\label{fig:BackgroundSubtraction2}
\end{figure}

To confirm that the observed decreasing signal (see main text Fig. 3) originates from the degradation of the nitroxide radicals rather than the NV centers, we conduct the experiments shown in Fig.~\ref{fig:BackgroundSubtraction2}. 

Figure \ref{fig:BackgroundSubtraction2}A illustrates the subtracted spectra from a long-term amplitude modulation measurement. Each spectrum undergoes 6.25 hours of exposure to a $ 37 \text{ W}/\text{cm}^2 $ laser. After a total illumination time of 56.25 hours, the majority of the radicals are quenched, allowing the raw data from the last measurement to be used as the background baseline. The curves in Fig.~\ref{fig:BackgroundSubtraction2}A are obtained by subtracting this background baseline from the raw data collected during the preceding 50 hours of illumination, and then removing the remaining slanted baseline. The same data processing procedure is applied to Fig.~3 in the main text. It is evident that the contrast of the radical signal decreases significantly over the long-term measurement. The contrast decay can be fitted by an exponential function with a characteristic time of 22(4) hours, as demonstrated by the grey line in Fig.~\ref{fig:BackgroundSubtraction2}C. 

Simultaneously, we detect the signal from an artificially applied RF field. Fig.~\ref{fig:BackgroundSubtraction2}B shows the raw spectra obtained by sweeping $f_{\text{RF}}$ instead of $f_{\text{mod}}$, which eliminates the impact of background baseline. The profile of the double-peak structure with FWHM \~{} 2 MHz matches the intrinsic line shape of the ensemble NV centers, which can be fitted by a double-peak Lorentzian function. The contrast of the RF field remained stable(Fig.~\ref{fig:BackgroundSubtraction2}D), indicating that the signal decay in the main text Fig.~3 and Fig.~\ref{fig:BackgroundSubtraction2}A resulted from the degradation of the radical signal.

\clearpage
\subsection{Supplementary Note 6. Calibration of spectral baselines with a reference region}
\begin{figure}[h]
		\centering
		\includegraphics[width=0.8\textwidth]{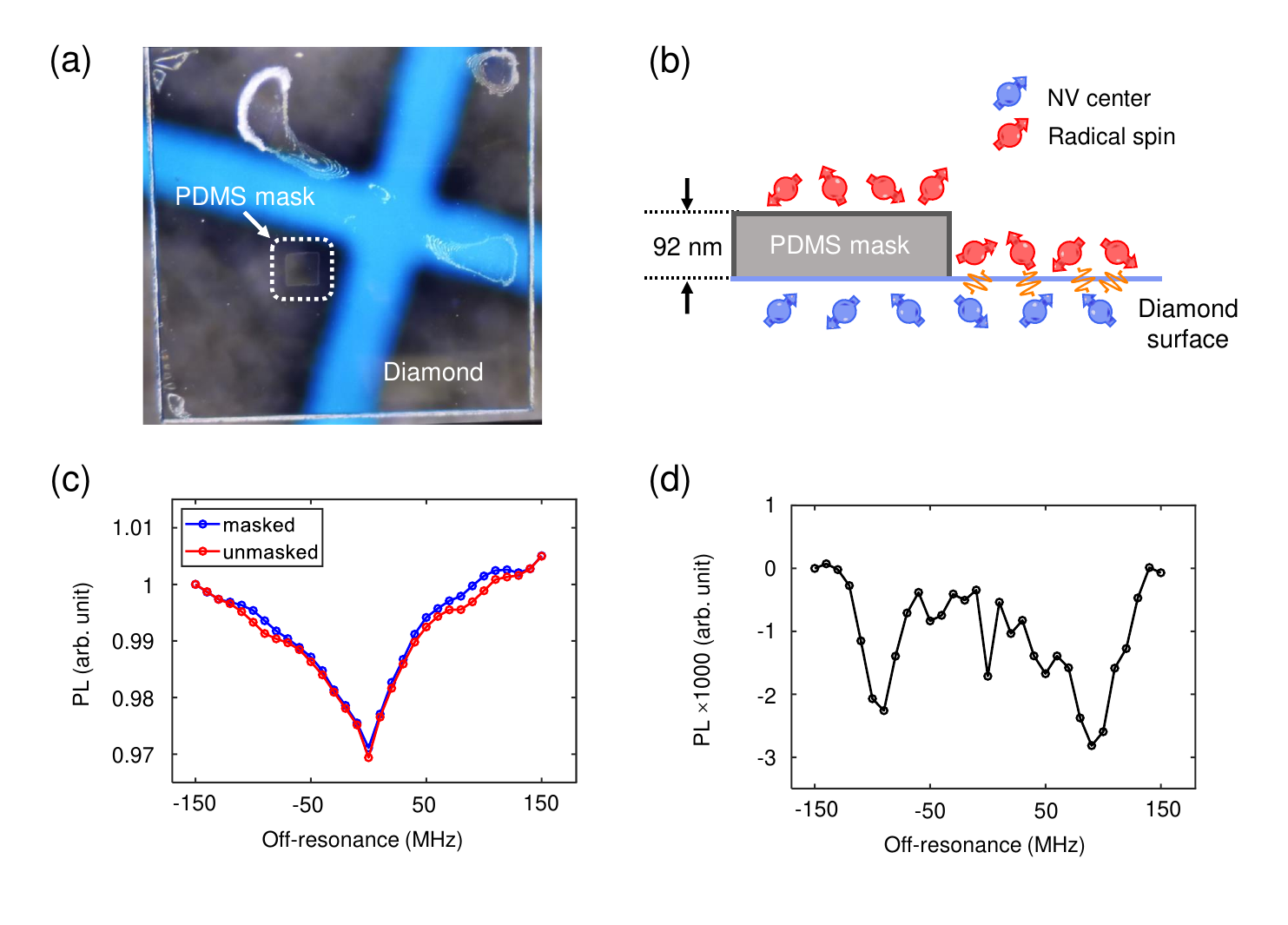}
		\caption{%
			\textbf{Calibration of spectral baselines with a masked region.}
			\revisen{(a) (b) Optical image and schematic diagram of the diamond with a patterned PDMS mask (thickness: 92 nm). Near-surface NV centers (blue arrows) detect the signals of radical spins (represented by red arrows) in the unmasked region, while no signal is detected in the masked region.
			(c) The raw data of the ‘off-resonance’ spectra of \textsuperscript{14}N-TEMPO methacrylate. Data from masked (blue) and unmasked (red) regions is acquired under identical experimental conditions ($\kappa = 0.3, \tau = 1$ ms).
			(d) The ‘off-resonance’ spectrum after background baseline subtraction. }
								}
		\label{fig:mask}
	\end{figure}

	\revisen{
	An alternative approach for calibrating the background baseline is to utilize a reference region without target spins. As illustrated in Fig.~\ref{fig:mask}(a,b), the diamond surface is patterned with a polydimethylsiloxane (PDMS) photoresist mask, followed by uniform deposition of PMMA containing \SI{100}{\milli\Molar} \textsuperscript{14}N-TEMPO methacrylate via spin-coating (4800 rpm). The PDMS thickness, measured using a step height profiler, is determined to be 92 nm, which exceeds the detection range of near-surface NV centers for radical spin detection. Consequently, NV centers beneath the mask record reference spectra corresponding to the background baseline, while those in unmasked regions detect spin-specific spectral features, as demonstrated in Fig.~\ref{fig:mask}(c). The measurement is conducted using the ‘off-resonance’ method \cite{Qin2023}, which shows no significant difference from the ‘amplitude-modulated’ method under current experimental conditions ($\kappa = 0.21$). By subtracting the masked reference data from the unmasked signal, the spectrum reveals two distinct peaks at ‘off-resonance frequency’ $f = \pm95$ MHz (Fig.~\ref{fig:mask}(d)), corresponding to the strongest resonance transition of \textsuperscript{14}N-TEMPO methacrylate (see SI Note 2).}
	
\FloatBarrier
\renewcommand\refname{Reference}

\bibliographystyle{naturemag}